\documentstyle[graphicx,times]{mn}

\newif\ifAMStwofonts
\ifoldfss
  \newcommand{\rmn}[1] {{\rm #1}}

  \ifCUPmtlplainloaded \else
    \NewTextAlphabet{textbfit} {cmbxti10} {}
    \NewTextAlphabet{textbfss} {cmssbx10} {}
    \NewMathAlphabet{mathbfit} {cmbxti10} {} % for math mode
    \NewMathAlphabet{mathbfss} {cmssbx10} {} %  "   "    "
  \fi
  \ifAMStwofonts
    \ifCUPmtlplainloaded \else
      \NewSymbolFont{upmath} {eurm10}
      \NewSymbolFont{AMSa} {msam10}
      \NewMathSymbol{\upi}     {0}{upmath}{19}
      \NewMathSymbol{\umu}     {0}{upmath}{16}
      \NewMathSymbol{\upartial}{0}{upmath}{40}
      \NewMathSymbol{\leqslant}{3}{AMSa}{36}
      \NewMathSymbol{\geqslant}{3}{AMSa}{3E}

       \let\le=\leqslant
       
    \fi
  \fi
\fi % End of OFSS

\ifnfssone
  \newmathalphabet{\mathit}
  \addtoversion{normal}{\mathit}{cmr}{m}{it}
  \addtoversion{bold}{\mathit}{cmr}{bx}{it}
  \newcommand{\rmn}[1] {\mathrm{#1}}

  \newmathalphabet{\mathbfit} % math mode version of \textbfit{..}
  \addtoversion{normal}{\mathbfit}{cmr}{bx}{it}
  \addtoversion{bold}{\mathbfit}{cmr}{bx}{it}
  \newmathalphabet{\mathbfss} % math mode version of \textbfss{..}
  \addtoversion{normal}{\mathbfss}{cmss}{bx}{n}
  \addtoversion{bold}{\mathbfss}{cmss}{bx}{n}
  \ifAMStwofonts
    \ifCUPmtlplainloaded \else
      \UseAMStwoboldmath
      \makeatletter
      \new@mathgroup\upmath@group
      \define@mathgroup\mv@normal\upmath@group{eur}{m}{n}
      \define@mathgroup\mv@bold\upmath@group{eur}{b}{n}
      \edef\UPM{\hexnumber\upmath@group}
      \new@mathgroup\amsa@group
      \define@mathgroup\mv@normal\amsa@group{msa}{m}{n}
      \define@mathgroup\mv@bold\amsa@group{msa}{m}{n}
      \edef\AMSa{\hexnumber\amsa@group}
      \makeatother
      \mathchardef\upi="0\UPM19
      \mathchardef\umu="0\UPM16
      \mathchardef\upartial="0\UPM40
      \mathchardef\leqslant="3\AMSa36
      \mathchardef\geqslant="3\AMSa3E

       \let\le=\leqslant

    \fi
  \fi
\fi % End of NFSS release 1

\ifnfsstwo
  \newcommand{\rmn}[1] {\mathrm{#1}}

  \DeclareMathAlphabet{\mathbfit}{OT1}{cmr}{bx}{it}
  \SetMathAlphabet\mathbfit{bold}{OT1}{cmr}{bx}{it}
  \DeclareMathAlphabet{\mathbfss}{OT1}{cmss}{bx}{n}
  \SetMathAlphabet\mathbfss{bold}{OT1}{cmss}{bx}{n}
  \ifAMStwofonts
    \ifCUPmtlplainloaded \else
      \DeclareSymbolFont{UPM}{U}{eur}{m}{n}
      \SetSymbolFont{UPM}{bold}{U}{eur}{b}{n}
      \DeclareSymbolFont{AMSa}{U}{msa}{m}{n}
      \DeclareMathSymbol{\upi}{0}{UPM}{"19}
      \DeclareMathSymbol{\umu}{0}{UPM}{"16}
      \DeclareMathSymbol{\upartial}{0}{UPM}{"40}
      \DeclareMathSymbol{\leqslant}{3}{AMSa}{"36}
      \DeclareMathSymbol{\geqslant}{3}{AMSa}{"3E}

       \let\le=\leqslant

    \fi
  \fi
\fi % End of NFSS release 2

\ifCUPmtlplainloaded \else
  \ifAMStwofonts \else % If no AMS fonts
    \def\upi{\pi}
    \def\umu{\mu}
    \def\upartial{\partial}
  \fi
\fi

\title[Structure of rich SMC clusters]
  {Surface brightness profiles and structural parameters for 10 rich stellar clusters in the Small Magellanic Cloud}
\author[A.~D.~Mackey \& G.~F.~Gilmore]
  {A.~D.~Mackey$^1$\thanks{E-mail: dmackey@ast.cam.ac.uk}
  and G.~F.~Gilmore$^1$\\
  $^1$Institute of Astronomy, University of Cambridge, Madingley Road,
  Cambridge CB3 0HA}
\date{Accepted --. Received --}
\pagerange{000--000}
\pubyear{2001}

\def\LaTeX{L\kern-.36em\raise.3ex\hbox{a}\kern-.15em
    T\kern-.1667em\lower.7ex\hbox{E}\kern-.125emX}

\begin{document}

\label{firstpage}

\maketitle

\begin{abstract}
As a follow up to our recent study of a large sample of LMC clusters
\cite{me}, we have conducted a similar study of the structures of ten
SMC clusters, using archival {\em Hubble Space Telescope} snapshot data.
We present surface brightness profiles for each cluster, and derive
structural parameters, including core radii and luminosity and mass
estimates, using exactly the same procedure as for the LMC sample.

Because of the small sample size, the SMC results are not as detailed 
as for the larger LMC sample. We do not observe any post core-collapse
clusters (although we did not expect to), and there is little evidence for
any double clusters in our sample. Nonetheless, despite the small sample
size, we show for the first time that the SMC clusters follow almost exactly the
trend in core radius with age observed for the LMC system, including
the apparent bifurcation at several hundred Myr. This further strengthens
our argument that this relationship represents true physical evolution
in these clusters, with some developing significantly expanded cores 
due to an as yet unidentified physical process. Additional data, both 
observational and from $N$-body simulations, is still required to 
clarify many issues.
\end{abstract}

\begin{keywords}
galaxies: star clusters -- Magellanic Clouds -- globular clusters: general -- stars: statistics
\end{keywords}

\section{Introduction}
We recently conducted a study of the structures of a large sample of 
rich star clusters in the Large Magellanic Cloud (LMC) (Mackey \& Gilmore
2002; hereafter Paper I), in which we compiled a pseudo-snapshot
data set from the {\em Hubble Space Telescope} ({\em HST}) archive
and used these observations to construct high resolution surface 
brightness profiles. From these profiles, we were able to obtain 
measurements of the structural parameters of each cluster, including 
their core radii, and total luminosities and masses. We also demonstrated
that these clusters followed a trend in core radius with age -- namely 
that the spread in core radius increases significantly as the clusters 
grow older, a result previously discussed by Elson and collaborators
\cite{efl,elson,ellson}. It seems likely that this trend reflects real
physical evolution of these clusters, as argued in Paper I, although
the mechanism by which the cores of some clusters expand during their
lifetimes while the cores of others do not, is as yet unidentified. 
We are currently carrying out $N$-body simulations to investigate 
in detail several physical processes which might drive core expansion
(e.g., Wilkinson et al., in prep.).

Having exhausted the suitable archival LMC data, we now turn our 
attention to the cluster system in the Small Magellanic Cloud (SMC). 
The SMC cluster system is similar to that of the LMC in that it contains
rich star clusters of masses comparable to Galactic globular clusters, 
and with ages spanning the range $10^{6}-10^{10}$ yr. The two systems
however, have dissimilar cluster formation histories and age-metallicity
relationships (see e.g., Rich et al. \shortcite{rsz}, for a brief 
review). The number of SMC clusters is also far fewer than the 
number of clusters in the LMC, and the SMC system
has been far less extensively studied. It is not surprising then that
there are few available surface brightness profiles for SMC clusters,
and (prior to the present study) it was not known whether the SMC cluster
system followed the core radius vs. age relationship observed for LMC 
clusters. In fact, we were only able to locate in the literature three 
low resolution density profile studies from photographic plates -- those
of Kontizas, Danezis \& Kontizas \shortcite{kdk}, Kontizas \& Kontizas 
\shortcite{kk}, and Kontizas, Theodossiou \& Kontizas \shortcite{ktk}. 

It therefore seemed very worthwhile to extract what archival {\em HST} 
frames we could locate, and reduce these using the same procedure we had
applied to the LMC sample, so that we now have two directly comparable 
and entirely homogeneous data sets. We describe these data in
Section \ref{sample} and briefly reiterate the reduction process in
Section \ref{reduction} -- this process and the problems and 
uncertainties associated with it have been described in great detail in
Paper I. We present the surface brightness profiles and the derivation
of key structural parameters in Section \ref{results}, and a discussion
of these results in the context of the core radius vs. age trend in 
Section \ref{discussion}. The results from
the present study (Tables \ref{data}, \ref{ages}, \ref{params}, and
\ref{luminmass}) together with the surface brightness profiles, are
available on-line at 
{\em http://www.ast.cam.ac.uk/STELLARPOPS/SMC\_clusters/}.

\section{The cluster sample}
\label{sample}
\subsection{Observations}
Just as for the LMC cluster sample, the observational basis of this 
project is archival {\em HST} data, again from {\em HST} project 5475 
which was a Wide Field Planetary Camera 2 (WFPC2) snapshot survey of 
Magellanic Cloud clusters. The data consists of two WFPC2 exposures per 
field, with the F450W and F555W filters respectively. Exposure times
span the ranges 80-600 s in F450W and 40-300 s in F555W, with the 
observation dates between 1994 January 26 and 1995 January 21. Around 
half of the SMC cluster exposures were taken before the WFPC2 cool-down 
of 1994 April 23.

Although thirteen clusters were targeted by this survey, we accepted 
only ten for reduction. The observations of NGC 422 were of poor data 
quality, and those for NGC 465 did not have a rich cluster suitable
for the construction of a surface brightness profile in the field of 
view. As noted by Crowl et al. \shortcite{crowl}, the observation 
labelled as NGC 419 is in fact a duplicate observation of NGC 411, at
a different roll angle. Exclusion of these three observations left a 
homogeneous sample of ten, which are listed in Table \ref{data} along 
with their observation details. Unlike with the LMC sample, we were not 
able to extend the current sample by including additional archival 
data -- no suitable observations could be located.

We note however, the existence of additional imaging of NGC 121 and NGC 330
in the {\em HST} archive. In the interests of maintaining consistency 
with our LMC data and our established reduction procedure, we chose not to use 
these observations. Paper I contains a full discussion of the existence of
extra archival data, and the reasons for its non-use.

\begin{table*}
\begin{minipage}{162mm}
\caption{Cluster list and observation details.}
\begin{tabular}{@{}lccccccccccc}
\hline \hline
Cluster & Program & & \multicolumn{4}{c}{Principal Frame} & & \multicolumn{4}{c}{Secondary Frame} \\
\cline{4-7}  \cline{9-12} \vspace{-3mm} \\
Name & ID & & Filter & Data set & Date & Time (s) & & Filter & Data set & Date & Time (s) \\
\hline
NGC121 & 5475 & & F555W & u26m0102t & 26/01/1994 & 300 & & F450W & u26m0101t & 26/01/1994 & 600 \\
NGC152 & 5475 & & F555W & u26m0702t & 26/09/1994 & 160 & & F450W & u26m0701t & 26/09/1994 & 300 \\
NGC176 & 5475 & & F555W & u26m0802t & 21/01/1995 & 100 & & F450W & u26m0801t & 21/01/1995 & 200 \\
NGC330 & 5475 & & F555W & u26m0b02t & 27/01/1994 & 40 & & F450W & u26m0b01t & 27/01/1994 & 80 \\
NGC339 & 5475 & & F555W & u26m0202t & 07/04/1994 & 200 & & F450W & u26m0201t & 07/04/1994 & 400 \\
NGC361 & 5475 & & F555W & u26m0602t & 03/04/1994 & 160 & & F450W & u26m0601t & 03/04/1994 & 300 \\
NGC411 & 5475 & & F555W & u26m0402t & 24/05/1994 & 200 & & F450W & u26m0401t & 24/05/1994 & 400 \\
NGC416 & 5475 & & F555W & u26m0502t & 06/02/1994 & 200 & & F450W & u26m0501t & 06/02/1994 & 400 \\
NGC458 & 5475 & & F555W & u26m0a02t & 03/02/1994 & 40 & & F450W & u26m0a01t & 03/02/1994 & 80 \\
KRON3 & 5475 & & F555W & u26m0g02t & 27/05/1994 & 300 & & F450W & u26m0g01t & 27/05/1994 & 600 \\
\hline
\label{data}
\end{tabular}
\end{minipage}
\end{table*}

\begin{table*}
\begin{minipage}{151mm}
\caption{Literature nomenclature, position, age and metallicity data for the cluster sample.}
\begin{tabular}{@{}llccccccc}
\hline \hline
Principal & Alternative & \multicolumn{3}{c}{Position (J2000.0)} & $\log\tau$ & Age & Metallicity & Met. \vspace{0.5mm} \\
Name & Names & $\alpha$ & $\delta$ & $R_{opt}$ $(\degr)^{a}$ & (yr) & Ref. & $[$Fe$/$H$]$ & Ref. \\
\hline
NGC121 & KRON2, L10 & $00^{h}26^{m}49^{s}$ & $-71\degr 32\arcmin 10\arcsec$ & $2.43$ & $10.08 \pm 0.05$ & $7$ & $-1.71 \pm 0.10$ & $7$ \vspace{0.2mm} \\
NGC152 & KRON10, L15 & $00^{h}32^{m}56^{s}$ & $-73\degr 06\arcmin 59\arcsec$ & $1.47$ & $9.15^{+0.06}_{-0.07}$ & $3$ & $-0.94 \pm 0.15$ & $3$ \vspace{0.2mm} \\
NGC176 & KRON12, L16 & $00^{h}35^{m}59^{s}$ & $-73\degr 09\arcmin 57\arcsec$ & $1.26$ & $8.30^{+0.30}_{-0.30}$ & $6,8$ & $\sim -0.6$ & $*$ \vspace{0.2mm} \\
NGC330 & KRON35, L54 & $00^{h}56^{m}20^{s}$ & $-72\degr 27\arcmin 44\arcsec$ & $0.46$ & $7.40^{+0.20}_{-0.40}$ & $2$,$4$ & $-0.82 \pm 0.11$ & $5$ \vspace{0.2mm} \\
NGC339 & KRON36, L59 & $00^{h}57^{m}45^{s}$ & $-74\degr 28\arcmin 21\arcsec$ & $1.68$ & $9.80^{+0.08}_{-0.10}$ & $7$ & $-1.50 \pm 0.14$ & $7$ \vspace{0.2mm} \\
NGC361 & KRON46, L67 & $01^{h}02^{m}11^{s}$ & $-71\degr 36\arcmin 25\arcsec$ & $1.43$ & $9.91^{+0.06}_{-0.07}$ & $7$ & $-1.45 \pm 0.11$ & $7$ \vspace{0.2mm} \\
NGC411 & KRON60, L82 & $01^{h}07^{m}56^{s}$ & $-71\degr 46\arcmin 09\arcsec$ & $1.59$ & $9.15^{+0.06}_{-0.07}$ & $1$,$3$ & $-0.68 \pm 0.07$ & $1$,$3$ \vspace{0.2mm} \\
NGC416 & KRON59, L83 & $01^{h}07^{m}58^{s}$ & $-72\degr 21\arcmin 25\arcsec$ & $1.25$ & $9.84^{+0.06}_{-0.08}$ & $7$ & $-1.44 \pm 0.12$ & $7$ \vspace{0.2mm} \\
NGC458 & KRON69, L96 & $01^{h}14^{m}54^{s}$ & $-71\degr 32\arcmin 58\arcsec$ & $2.17$ & $8.30^{+0.18}_{-0.30}$ & $4$ & $-0.23^{+0.1}_{-0.4}$ & $4$ \vspace{0.2mm} \\
KRON3 & L8 & $00^{h}24^{m}46^{s}$ & $-72\degr 47\arcmin 37\arcsec$ & $2.07$ & $9.78^{+0.09}_{-0.11}$ & $7$ & $-1.16 \pm 0.09$ & $7$ \vspace{0.2mm} \\
\hline
\label{ages}
\end{tabular}
\medskip
\\
Reference list: 1. Alves \& Sarajedini \shortcite{alves}; 2. Chiosi et al. \shortcite{chiosi}; 3. Crowl et al. \shortcite{crowl}; 4. Da Costa \& Hatzidimitriou \shortcite{gdc}; 5. Hill \shortcite{hill}; 6. Hodge \& Flower \shortcite{hodgeflower}; 7. Mighell et al. \shortcite{mighell}; 8. This paper.\\
$^{*}$ Calculated metallicity, as described in the text.\\
$^{a}$ Relative to the optical centre of the SMC, at $\alpha = 00^{h}52^{m}45^{s}$, $\delta = -72\degr 49\arcmin 43\arcsec$ (J2000.0)\ \cite{westerbook}
\end{minipage}
\end{table*}

\subsection{Literature data}
We have compiled literature data for the identifiers, positions, ages
and metallicities of the ten clusters in the present sample. These data
are presented in Table \ref{ages}. Like for 
the previous LMC sample, we have tried to make this compilation as 
homogeneous as possible, which means taking the results of larger scale 
studies where possible, but without sacrificing the quality of the data 
by neglecting other high resolution or high accuracy measurements. As 
with the compilation in Paper I, this compilation is not intended to be 
an exhaustive survey of the available literature; rather it merely 
provides a consistent set of age and metallicity estimates as a 
reference point for this and future work. 

\subsubsection{Cluster names and positions}
We have taken the most common identifiers for each cluster from the
Simbad Astronomical Database ({\em http://simbad.u-strasbg.fr/}).
In all but one case the principal designation is an NGC number, with the
other labels (in parenthesis) being drawn from the catalogues of 
Kron \shortcite{kron} (Kron) and Lindsay \shortcite{lindsay} 
(Lindsay; abbreviated to `L' in tables and figures). Positions are
taken from Welch \shortcite{welch}, precessed to J2000.0, except that 
for NGC 121 which is from Simbad. Although we later derive considerably
more precise positions, none of these are significantly different from
the literature values described here. We have also listed the projected 
angular distance $R_{opt}$ to the optical centre of the SMC at 
$\alpha = 00^{h}52^{m}45^{s}$, $\delta = -72\degr 49\arcmin 43\arcsec$ 
(J2000.0)\ \cite{westerbook}. Because the SMC has a complex disrupted
structure, its rotation centre (and rotation curve) are not easily 
derived and remain the subject of debate \cite{westerbook}. We therefore
do not quote projected angular distances from a rotation centre here,
but consider this further in Section \ref{discussion}, below.

\subsubsection{Cluster ages}
\label{clusterages}
The SMC cluster system is less extensively studied than the LMC system,
and age determinations based on colour-magnitude diagrams (CMDs) from 
CCD measurements are correspondingly sparse. We have adopted the majority
of our ages from the analysis of Mighell, Sarajedini \& French 
\shortcite{mighell}, who determine CMDs for five of the clusters in
the present sample (NGC 121, NGC 339, NGC 361, NGC 416, and Kron 3) from exactly 
the same archival {\em HST} snapshot data.
We take the ages of NGC 152 and NGC 411 from Crowl et al. 
\shortcite{crowl} and Alves \& Sarajedini \shortcite{alves} respectively,
with these two studies using similar analysis techniques to those of
Mighell et al. Rich et al. \shortcite{rich} have published
a separate analysis of the same archival data using a different age
determination technique, and find NGC 152 and NGC 411 to be coeval
($\tau = 2 \pm 0.5$ Gyr) and NGC 339, NGC 361, NGC 416 and Kron 3 to be 
coeval ($\tau = 8 \pm 2$ Gyr). Both results are consistent within the 
errors with the ages adopted from the three studies listed above. 
For NGC 330, we take an age estimate from the study of Chiosi et al.
\shortcite{chiosi}; see also the discussion of Da Costa \&
Hatzidimitriou \shortcite{gdc}. Da Costa \& Hatzidimitriou also discuss 
literature estimates for the age of NGC 458 and we take the average
of the two values mentioned in this discussion. Finally, for NGC 176
we resort to the photographic plate photometry of Hodge \& Flower
\shortcite{hodgeflower} who suggest an age of $\sim 0.4$ Gyr. However,
our unpublished CMD suggests an age much more similar to that of NGC 458
or younger, so we adopt $\tau \sim 0.2$ Gyr, but with large error-bars to
cover both possibilities.

\subsubsection{Cluster metallicities}
Literature estimates for SMC cluster metallicities are also few and
far between. To maintain consistency, we have adopted the cluster 
metallicities determined in conjunction with their ages by Mighell 
et al. \shortcite{mighell} for NGC 121, NGC 339, NGC 361, NGC 416, and
Kron 3, by Alves \& Sarajedini \shortcite{alves} for NGC 411, and by
Crowl et al. \shortcite{crowl} for NGC 152. In each of these cases,
the uncertainties we quote are those formally derived by the authors,
and represent the uncertainties from both their data and their fitting
procedures. Da Costa \& Hatzidimitriou \shortcite{gdc} have measured 
spectroscopic abundances based on the
calcium triplet for three of these clusters -- NGC 121, NGC 339, and
Kron 3. Their age-corrected results show good consistency with the
values adopted here for these three clusters. For NGC 330, we take
the metallicity determined from the high resolution spectroscopy of
Hill \shortcite{hill}, and for NGC 458 the metallicity estimated from
the literature by Da Costa \& Hatzidimitriou. Finally, for NGC 176 we
could not locate any suitable metallicity measurements, so we have used
the SMC age-metallicity relationship presented by Da Costa \& 
Hatzidimitriou to estimate [Fe/H] $\sim -0.6$. This final
metallicity is not intended as anything more than a crude estimate, and
we adopt it solely for the purpose of estimating mass to light ratios
later in the analysis. In Paper I we showed that these calculations are
relatively insensitive to the selected metallicity, so we are confident
in using the metallicity for NGC 176 in this case. Any subsequent use
must be carefully judged on the nature of the calculation at hand.

\section{Photometry and surface brightness profiles}
\label{reduction}
The data reduction, photometry, and construction of surface brightness
profiles followed exactly the procedures outlined in Paper I, and for a
detailed description we refer the reader to this paper. For completeness,
we provide a much-abbreviated summary below.

As part of the retrieval process, all archival exposures are reduced 
according to the standard {\em HST} pipeline, using the latest available
calibrations. We used the HSTphot software package \cite{hstphot}, and
in particular the {\em multiphot} routine, to make our photometric
measurements from the reduced frames. HSTphot is tailored for WFPC2
observations, accounting in detail for the severely under-sampled PC/WFC
point spread functions (PSFs) and the four chip structure of the camera.
The images were first cleaned\footnote{i.e., the masking of bad regions 
and pixels, a first attempt at removing cosmic rays and hot and 
saturated pixels, and the robust determination of a background image.} 
using the packages provided with HSTphot,
and then the photometry performed using HSTphot in PSF fitting mode,
with a minimum detection threshold of $3\sigma$ above the local 
background. Parameters derived during the PSF fitting routines, such as
the object classification, sharpness parameter, and goodness-of-fit 
parameter $\chi$ helped clean the photometry of spurious detections and
non-stellar objects. We selected only stellar classified objects with
sharpnesses in the interval $[-0.6,+0.6]$ and $\chi \le 3.5$, which 
provided a perfectly adequate selection for the purposes of constructing
surface brightness profiles. Photometry for these selected objects
was corrected for geometric distortion, filter-dependent plate scale
changes, and the CCD 34th row errors. Finally, charge-transfer efficiency
(CTE) effects were corrected using the calibration of Dolphin 
\shortcite{wfpc}, PSF residuals accounted for, and the final 
measurements corrected to an aperture of $0\farcs5$ and the zero-points
of Dolphin \shortcite{wfpc}.

For each cluster, the selections of measured stars were used to 
independently construct two surface brightness profiles -- one for each 
colour. Comparison of each pair of profiles provided a good consistency
check. The difficulties associated with constructing profiles from
WFPC2 observations are outlined in Paper I; here we simply summarize
our procedure. First, the chip and pixel coordinates for each star are 
converted to corrected pixel coordinates relative to the WFC2 chip, using
the {\sc iraf}\ {\sc stsdas} task {\sc metric}. This procedure does not
use any image header information, so the inaccuracies which sometimes
occur in image headers are circumvented. Next, the centre of each 
cluster (or more correctly, the surface brightness peak of a cluster) 
is located using a simple Monte Carlo style algorithm. Accuracy in the 
centre determination is vital,
since poor centering will tend to artificially flatten a surface 
brightness profile, which will cause systematic errors in the 
determination of structural parameters. Our algorithm is repeatable
typically to $\pm 10$ WFC pixels, or approximately $\pm 1\arcsec$. 

Four sets of annuli are then generated about the centre. Two sets are
of narrow width ($1\farcs5$ and $2\arcsec$, extending to $\sim 20\arcsec$
and $\sim 30\arcsec$ respectively) and are designed for sampling of the
core, while the other two sets are wider ($3\arcsec$ and $4\arcsec$, 
both extending as far as possible) and are designed for sampling the 
outer regions. The surface brightness $\mu_{i}$ of the $i$-th annulus 
in a set is found simply by counting the stars which fall in that 
annulus, so that
\begin{equation}
\mu_{i} = \frac{A_{i}}{\pi (b^{2}_{i} - a^{2}_{i})} \sum_{j=1}^{N_{s}} C_{j} F_{j}   
\label{sb}
\end{equation}
where $b_{i}$ and $a_{i}$ are the outer and inner radii of the annulus 
respectively, $N_{s}$ is the number of stars in the annulus, and $F_{j}$
is the flux of the $j$-th star. The factors $A_{i}$ and $C_{j}$ are
the area correction for an annulus and the completeness correction
for a star respectively, and must be determined before the annulus
is constructed. 

The area correction for an annulus is designed to ensure the surface
brightness is normalized to that for a complete annulus -- this is 
necessary because the shape of the WFPC2 camera and the 
arbitrary cluster centering and observation roll-angle mean that
most annuli about the cluster centre are not fully imaged. If this factor
was not accounted for, serious artificial fluctuations would evidently
be introduced into each profile. Because of the complicated observation 
geometries, analytic determination of each area correction is overly
difficult; rather, we determined them numerically, as outlined in Paper
I. Annuli with $A_{i} > 3$ are not used -- this limits the maximum
profile radius to $\sim 75-80\arcsec$, depending on the position of
the cluster centre on the camera.

The completeness correction for a star is designed to account for those
stars missed by the detection software due to faintness and crowding. 
We used the artificial star routine attached to {\em multiphot} in the
manner described in Paper I, to develop a stellar completeness function 
which is dependent on chip and pixel coordinates, magnitude, and colour.
This function is in the form of a look-up table, so the appropriate
correction for the $j$-th star may be easily located according to the
star's properties. Stars with $C_{j} > 4$ were discarded to avoid 
introducing large random errors. Unlike several of the clusters in the
larger LMC sample, none of the clusters in the present sample suffered
from serious crowding or saturation, so the completeness corrections
were usually small, and the positional binning resolution described
in Paper I was always perfectly adequate. In addition, there was never 
a need for extra data from short exposures (see Paper I) to be introduced.

Just as with most of the LMC clusters in Paper I, saturated stars were 
also not a problem. Only one of the clusters in the present sample -- the
very young cluster NGC 330 -- had significant numbers of saturated stars
($\ga 10$) within approximately two core radii. This cluster appears
similar to the young LMC clusters NGC 1805 and NGC 1818, which suffered
similar numbers of saturated stars in the measurements made in Paper I.
In the LMC case, we demonstrated that it made sense to leave such stars
(predominantly very short lived giants and upper main sequence stars)
out of the surface brightness profile construction, in order to measure
properly the underlying stellar distribution. The case of NGC 330 is
exactly similar and we are therefore confident in leaving the 
saturated stars from this cluster out of any further calculations.

The internal errors $\sigma_{i}$ for each annulus were determined
initially according to the method outlined by Djorgovski \shortcite{djor}
-- that is, each annulus was divided into eight and the standard 
deviation of the segmental surface brightnesses taken to be the error.
As described in Paper I, this worked well in the inner regions of a 
cluster but broke down in the outer regions, badly underestimating the 
scatter. Therefore, after the background subtraction but before the 
final model fit (see below), we calculated the Poisson errors for each 
annulus, and in the cases where these were significantly larger than the 
original errors we substituted the new errors instead. For some clusters, 
particularly those of low density, both the Poisson 
and sector errors are larger than the RMS point-to-point scatter. This is a 
consequence of the calculation technique and is addressed at length in
Paper I (Section 4.2.5).

EFF models (after Elson, Fall \& Freeman \shortcite {eff}) 
of the form
\begin{equation}
\mu(r) = \mu_{0} \left(1 + \frac{r^{2}}{a^{2}} \right)^{-\frac{\gamma}{2}}
\label{ep}
\end{equation}
were fit to each profile, where $\mu_{0}$ is the central surface 
brightness, $a$ is a measure of the core radius and $\gamma$ is the 
power-law slope at large radii. These profiles are essentially the 
same as the family of empirical King \shortcite{king} models, without
tidal truncation. The parameter $a$ is related to
the usual King core radius $r_{c}$ by
\begin{equation}
r_{c} = a(2^{2 / \gamma} - 1)^{1 / 2}
\label{rc}
\end{equation}
provided the tidal cut-off $r_{t} >> r_{c}$; a safe assumption for most 
of the clusters in this sample, given the relatively weak tidal field of
the SMC.

\begin{table*}
\begin{minipage}{164mm}
\caption{Structural parameters for the cluster sample derived from the best-fitting F555W EFF profiles.}
\begin{tabular}{@{}lcccccccc}
\hline \hline
Cluster & \multicolumn{2}{c}{Centre (J2000.0)$^{a}$} & $\mu_{555}(0)^{b}$ & $a$ & $\gamma$ & $r_{c}$ & $r_{c}$ & $r_{m}$ \vspace{0.5mm} \\
Name & $\alpha$ & $\delta$ & & $(\arcsec)$ & & $(\arcsec)$ & (pc)$^{c}$ & $(\arcsec)$ \\
\hline
NGC121 & $00^{h}26^{m}48\fs8$ & $-71\degr 32\arcmin 12\arcsec$ & $18.50 \pm 0.06$ & $12.97 \pm 0.67$ & $3.17 \pm 0.08$ & $9.61 \pm 0.35$ & $2.81 \pm 0.10$ & 76 \\
NGC152 & $00^{h}32^{m}55\fs7$ & $-73\degr 07\arcmin 01\arcsec$ & $20.99 \pm 0.04$ & $20.24 \pm 3.37$ & $2.07 \pm 0.26$ & $19.74 \pm 1.45$ & $5.77 \pm 0.42$ & 76 \\
NGC176 & $00^{h}35^{m}59\fs2$ & $-73\degr 09\arcmin 59\arcsec$ & $20.35 \pm 0.08$ & $10.21 \pm 1.01$ & $2.64 \pm 0.14$ & $8.49 \pm 0.64$ & $2.48 \pm 0.19$ & 64 \\
NGC330 & $00^{h}56^{m}18\fs0$ & $-72\degr 27\arcmin 47\arcsec$ & $17.54 \pm 0.05$ & $10.58 \pm 0.82$ & $2.58 \pm 0.14$ & $8.93 \pm 0.40$ & $2.61 \pm 0.12$ & 72 \\
NGC339 & $00^{h}57^{m}47\fs3$ & $-74\degr 28\arcmin 22\arcsec$ & $20.93 \pm 0.06$ & $45.73 \pm 6.05$ & $5.21 \pm 0.99$ & $25.26 \pm 1.03$ & $7.38 \pm 0.30$ & 72 \\
NGC361 & $01^{h}02^{m}11\fs6$ & $-71\degr 36\arcmin 21\arcsec$ & $20.54 \pm 0.05$ & $18.49 \pm 1.73$ & $2.05 \pm 0.13$ & $18.17 \pm 0.97$ & $5.31 \pm 0.28$ & 76 \\
NGC411 & $01^{h}07^{m}56\fs0$ & $-71\degr 46\arcmin 03\arcsec$ & $19.31 \pm 0.04$ & $11.93 \pm 0.70$ & $2.72 \pm 0.11$ & $9.72 \pm 0.36$ & $2.84 \pm 0.11$ & 76 \\
NGC416 & $01^{h}07^{m}59\fs1$ & $-72\degr 21\arcmin 12\arcsec$ & $18.58 \pm 0.05$ & $14.42 \pm 1.19$ & $3.70 \pm 0.32$ & $9.73 \pm 0.35$ & $2.84 \pm 0.10$ & 72 \\
NGC458 & $01^{h}14^{m}52\fs1$ & $-71\degr 33\arcmin 05\arcsec$ & $18.99 \pm 0.08$ & $15.03 \pm 2.13$ & $3.22 \pm 0.34$ & $11.02 \pm 0.82$ & $3.22 \pm 0.24$ & 69 \\
KRON3 & $00^{h}24^{m}45\fs8$ & $-72\degr 47\arcmin 38\arcsec$ & $20.20 \pm 0.03$ & $26.97 \pm 1.47$ & $3.00 \pm 0.14$ & $20.67 \pm 0.62$ & $6.04 \pm 0.18$ & 76 \\
\hline
\end{tabular}
\medskip
\\
$^{a}$ We find our centering algorithm to be repeatable to approximately $\pm 1\arcsec$, notwithstanding image header inaccuracies. Given this precision, coordinates in $\delta$ are provided to the nearest arcsecond. Those in $\alpha$ are reported to the nearest tenth of a second, but the reader should bear in mind that at $\delta = -72\degr$, one second of RA corresponds to approximately five seconds of arc -- in other words, the uncertainty in $\alpha$ is approximately $\pm 0\fs2$.\\
$^{b}$ The $V_{555}$ magnitude of one square arcsecond at the centre of a given cluster.\\
$^{c}$ When converting to parsecs we assume an SMC distance modulus of $18.9$ which equates to a scale of $3.423$ arcsec pc$^{-1}$.
\label{params}
\end{minipage}
\end{table*}

The fitting of these profiles was achieved using weighted least-squares
minimization on an adaptive grid, as described in Paper I. Because
many of the clusters lie in well-populated regions of the SMC,
field star contamination needed to be corrected for. This was done
by initially fitting an EFF model to the central region of a profile,
where field contamination is not significant. This provided estimates
of the parameters $\mu_{0}$ and $r_{c}$, which were subsequently used to
fit, in the outer regions of the profile, a model of the form:
\[
\mu(r) \,\,\approx\,\, \mu_{0}\left(\frac{r}{a}\right)^{-\gamma} + \,\, \phi
\]
\begin{equation}
\hspace{7mm} \,\,\approx\,\, \mu_{0}\left(\frac{r}{r_{c}}\right)^{-\gamma} (2^{2/\gamma}-1)^{-\gamma/2} \,\, + \,\,\phi
\label{bgrfit}
\end{equation}
which is Eq. \ref{ep} with $r>>a$, a flat background contribution 
$\phi$, and Eq. \ref{rc} substituted. From this model we determined
$(\gamma,\phi)$, so that the field contamination level could be 
subtracted from each annulus in the profile. Finally, an EFF model 
was fit to the full subtracted profile and the structural parameters 
$(\mu_{0},\gamma,a)$ determined. This procedure includes
an assumption that the central parameters $(\mu_{0},r_{c})$ are
essentially independent of the background level. We demonstrated the
validity of this assumption, and indeed the entire background subtraction
process in Paper I. In particular, it was shown that no significant 
systematic errors were introduced into the structural parameter 
measurements and especially the determination of $\gamma$, which is 
very sensitive to the background level. 

Finally, uncertainties in the measured parameters $(\mu_{0},\gamma,a)$ 
were determined using a bootstrap method  (Press et al. 1992, p691)
with 1000 recursions, as described in Paper I.

\begin{table}
\caption{Previously published core radius measurements.}
\begin{tabular}{@{}lcccc}
\hline \hline
Cluster & $r_{c}$ (pc) & $r_{c}$ (pc) & $r_{c}$ (pc) & Ratio$^{c}$ \vspace{0.5mm} \\
 & (this paper) & (published)$^{a}$ & (converted)$^{b}$ & \\
\hline
NGC121 & $2.81 \pm 0.10$ & $4.5$  & $3.92$  & $0.72$  \\
NGC152 & $5.77 \pm 0.42$ & $13.2$ & $11.50$ & $0.50$  \\
NGC176 & $2.48 \pm 0.19$ & $1.0$  & $0.87$  & $2.85$  \\
NGC330 & $2.61 \pm 0.12$ & $1.9$  & $1.65$  & $1.58$  \\
NGC339 & $7.38 \pm 0.30$ & $0.7$  & $0.61$  & $12.10$ \\
NGC361 & $5.31 \pm 0.28$ & $3.2$  & $2.79$  & $1.90$  \\
NGC411 & $2.84 \pm 0.11$ & $2.5$  & $2.18$  & $1.30$  \\
NGC416 & $2.84 \pm 0.10$ & $1.2$  & $1.05$  & $2.70$  \\
NGC458 & $3.22 \pm 0.24$ & $2.7$  & $2.35$  & $1.37$  \\
KRON3  & $6.04 \pm 0.18$ & $2.4$  & $2.09$  & $2.89$  \\
\hline
\end{tabular}
\medskip
\\
$^{a}$ Based on an SMC distance modulus of 19.2. \\
$^{b}$ Converted to our distance scale, with an SMC distance modulus of 18.9. \\
$^{c}$ Column 2 divided by Column 4.
\label{comp}
\end{table}

\section{Results}
\label{results}
\subsection{Profiles and structural parameters}
\begin{figure*}
\begin{minipage}{175mm}
\includegraphics[width=87mm,height=79.5mm]{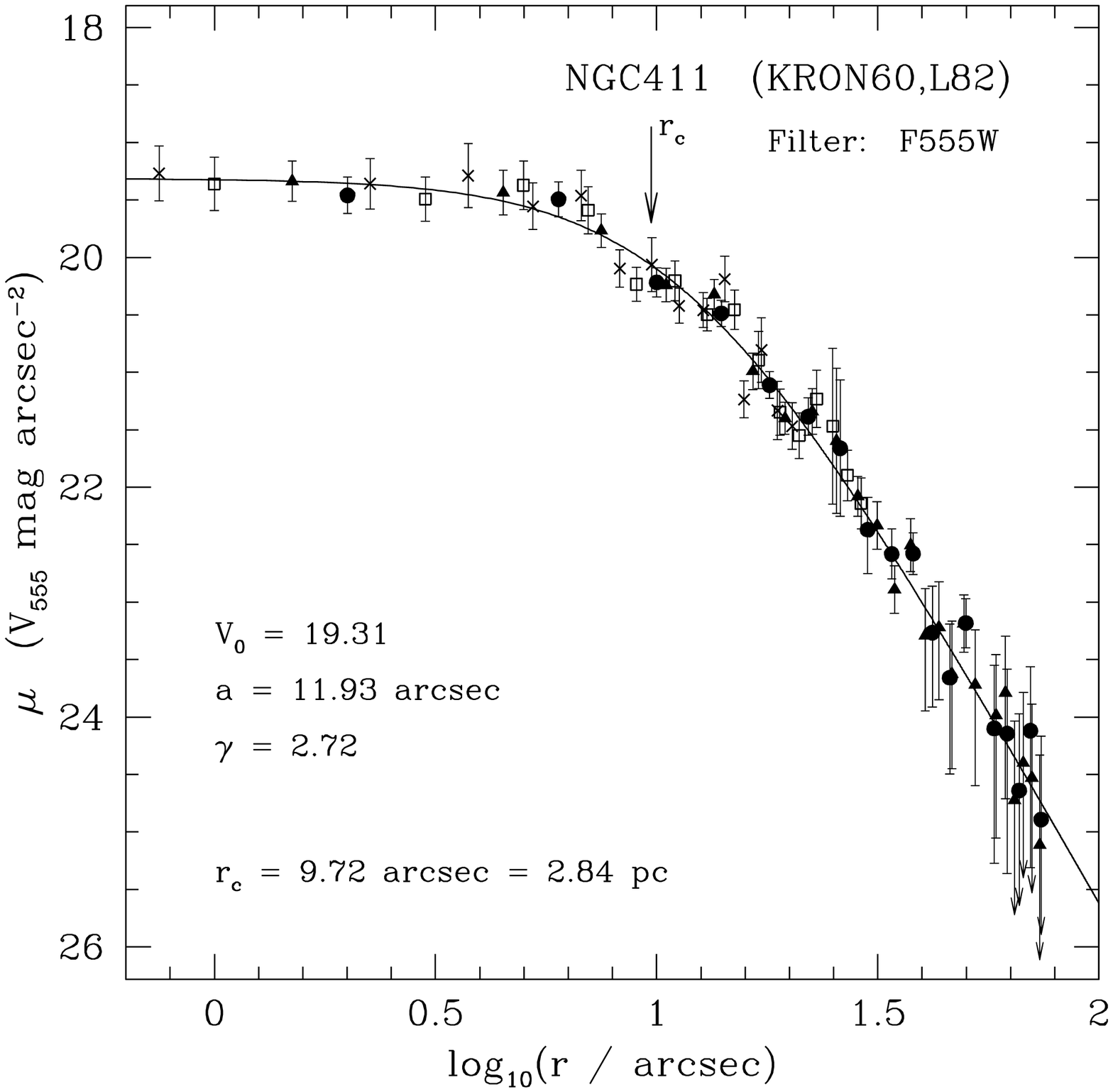}
\includegraphics[width=87mm,height=79.5mm]{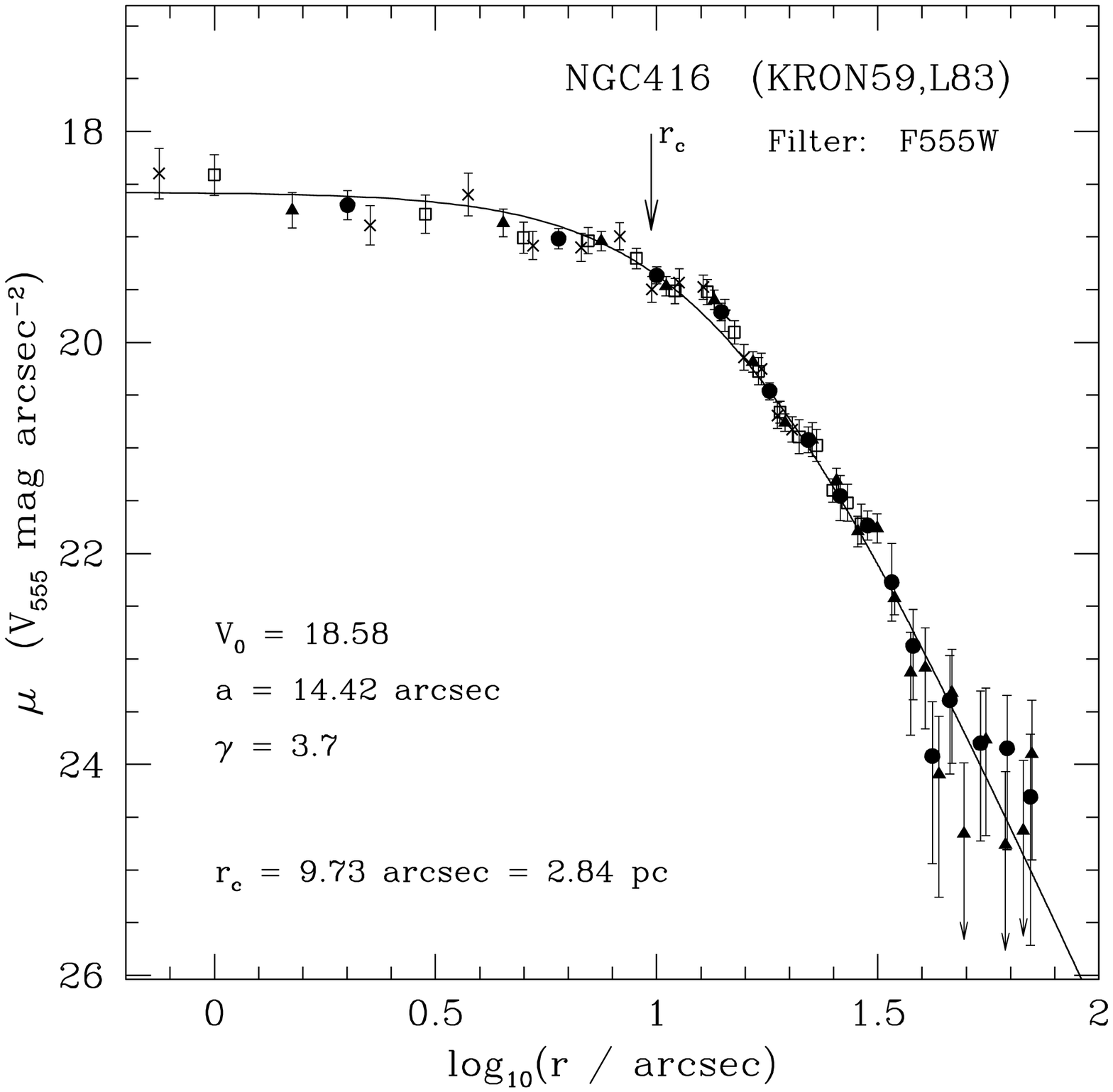}
\newline
\includegraphics[width=87mm,height=79.5mm]{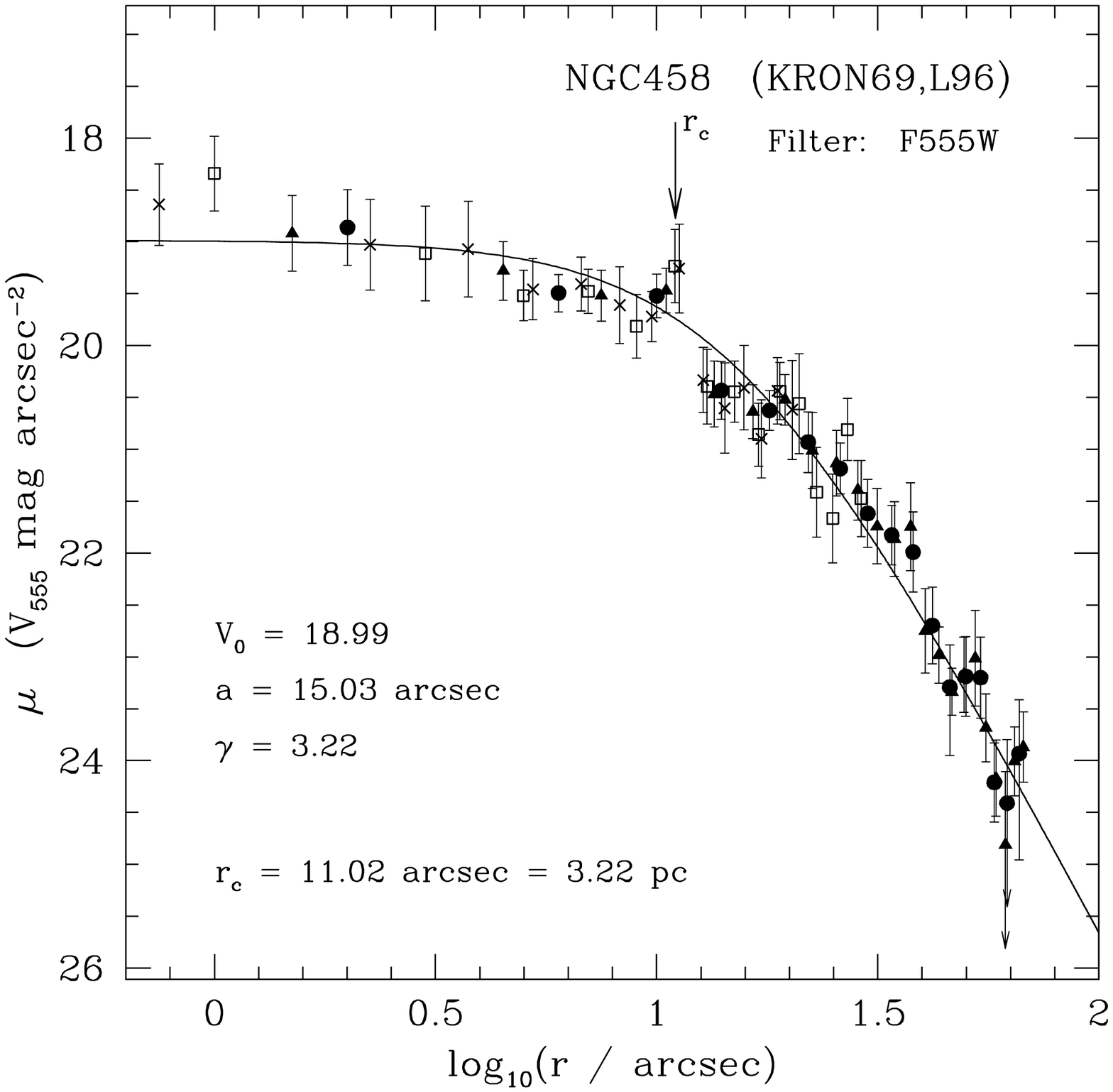}
\includegraphics[width=87mm,height=79.5mm]{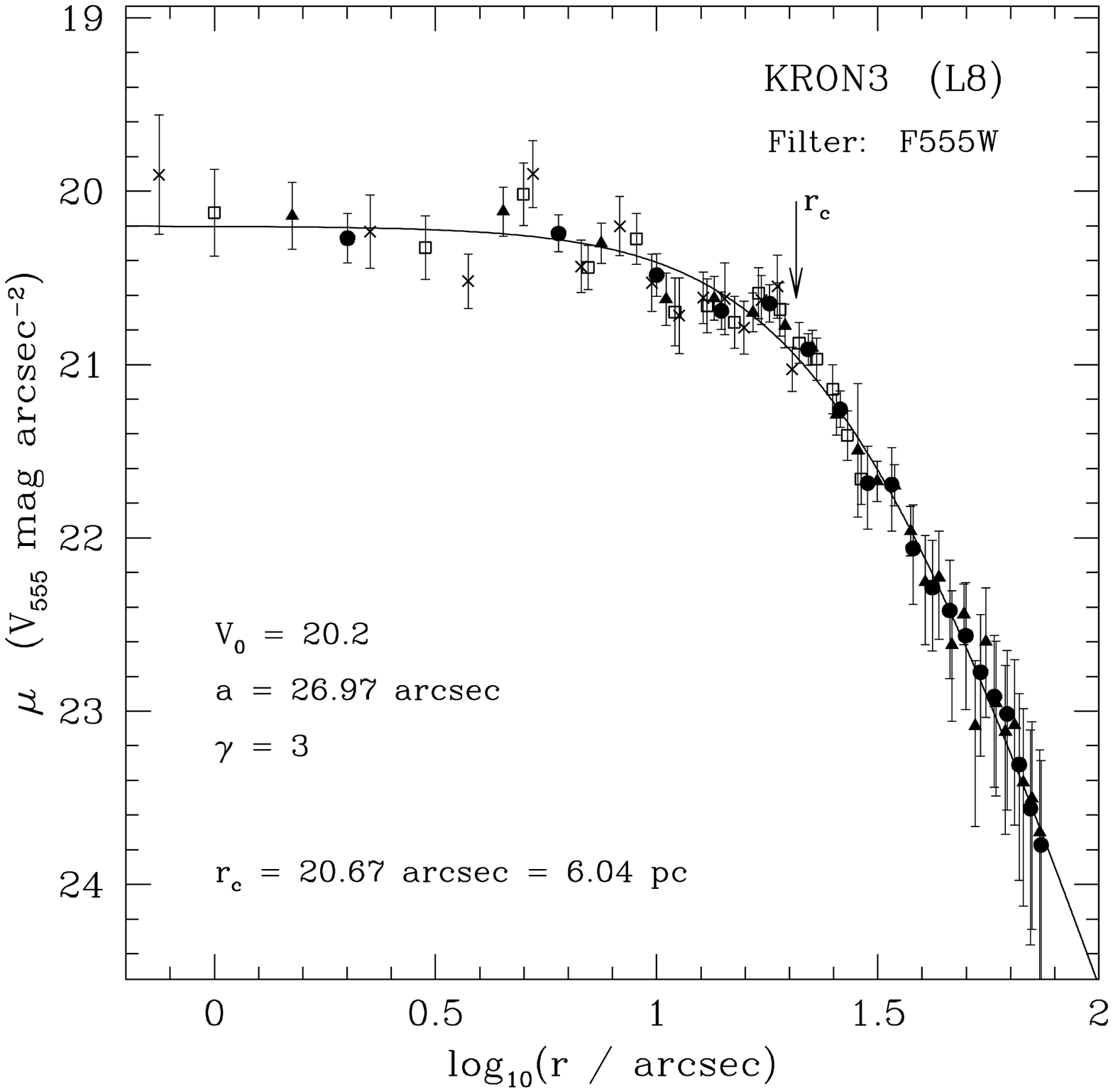}
\caption{Background-subtracted F555W surface brightness profiles for each of the 10 clusters in the sample. The four different annulus widths are marked with different point types: $1\farcs5$ width are crosses, $2\arcsec$ width are open squares, $3\arcsec$ width are filled triangles, and $4\arcsec$ width are filled circles. Error bars marked with down-pointing arrows fall below the bottom of their plot. The solid lines show the best-fit EFF profiles. For each cluster the core radius $r_{c}$ is indicated and the best-fit parameters listed. When converting to parsecs, we assume an SMC distance modulus of 18.9.}
\label{plots}
\end{minipage}
\end{figure*}

The background-subtracted F555W surface brightness profiles for each
of the 10 clusters are presented in Fig. \ref{plots}. To demonstrate the
high degree of consistency between them, we plot each of the four 
annulus sets on the same axes. For each cluster, the best fit 
EFF profile is also plotted, the core radius indicated, and the best 
fit parameters listed. These results are summarized in Table 
\ref{params} along with their corresponding errors, the calculated 
centre of each cluster and the maximum radial extent $r_{m}$ of each 
profile's measurements.

% SB profiles here... about 2 pages' worth.

\begin{figure*}
\begin{minipage}{175mm}
\includegraphics[width=87mm,height=79.5mm]{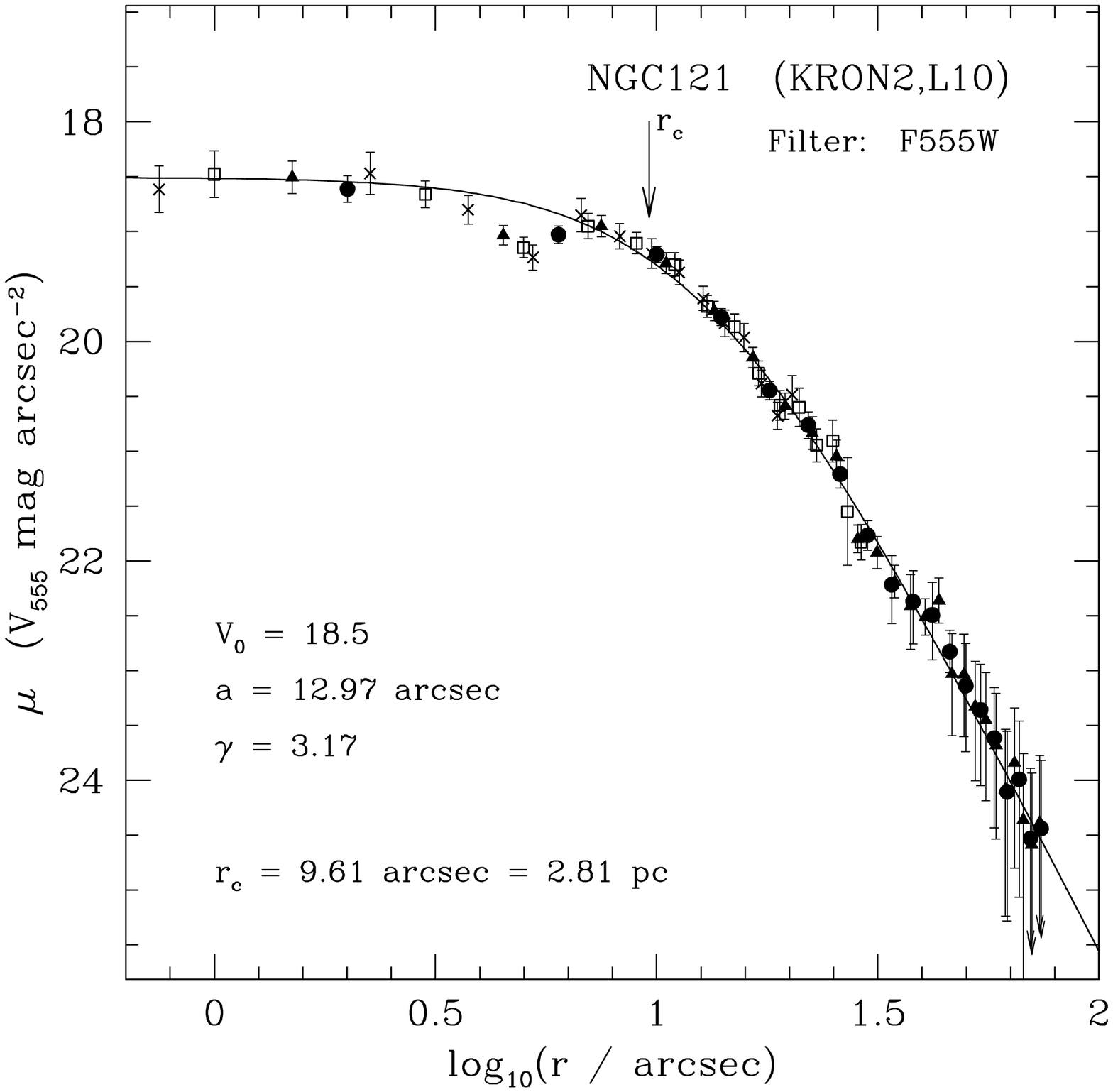}
\includegraphics[width=87mm,height=79.5mm]{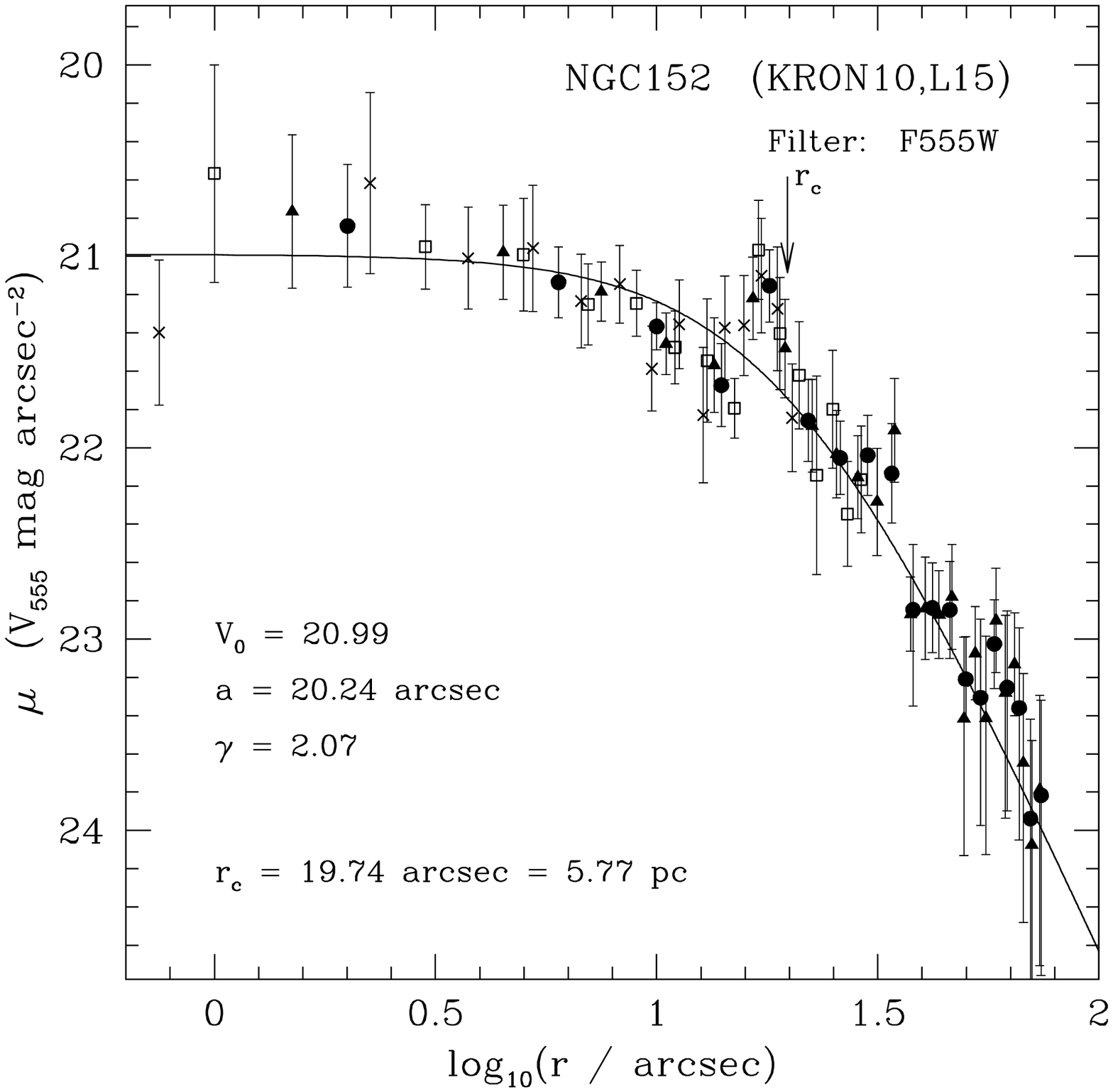}
\newline
\includegraphics[width=87mm,height=79.5mm]{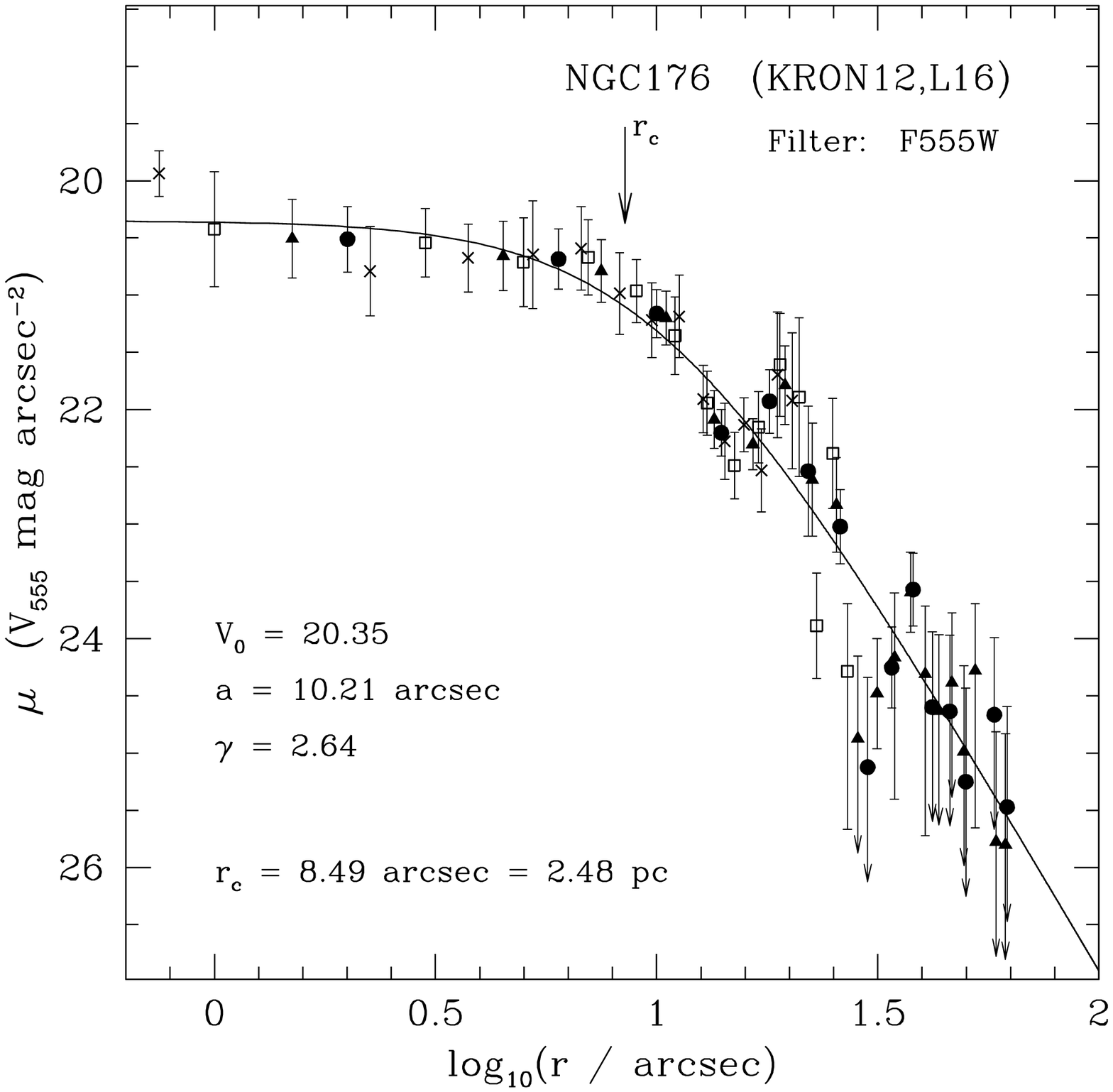}
\includegraphics[width=87mm,height=79.5mm]{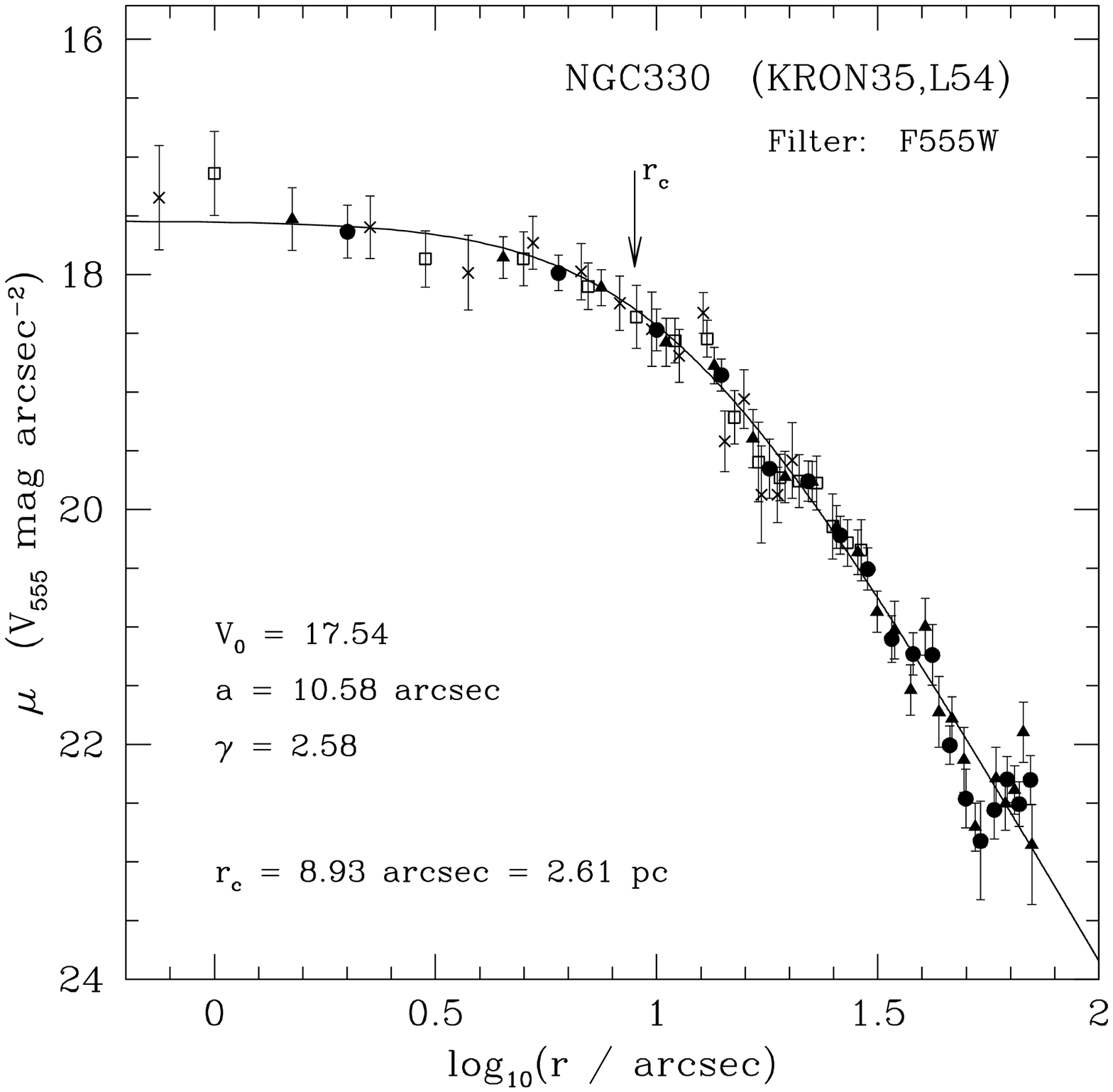}
\newline
\includegraphics[width=87mm,height=79.5mm]{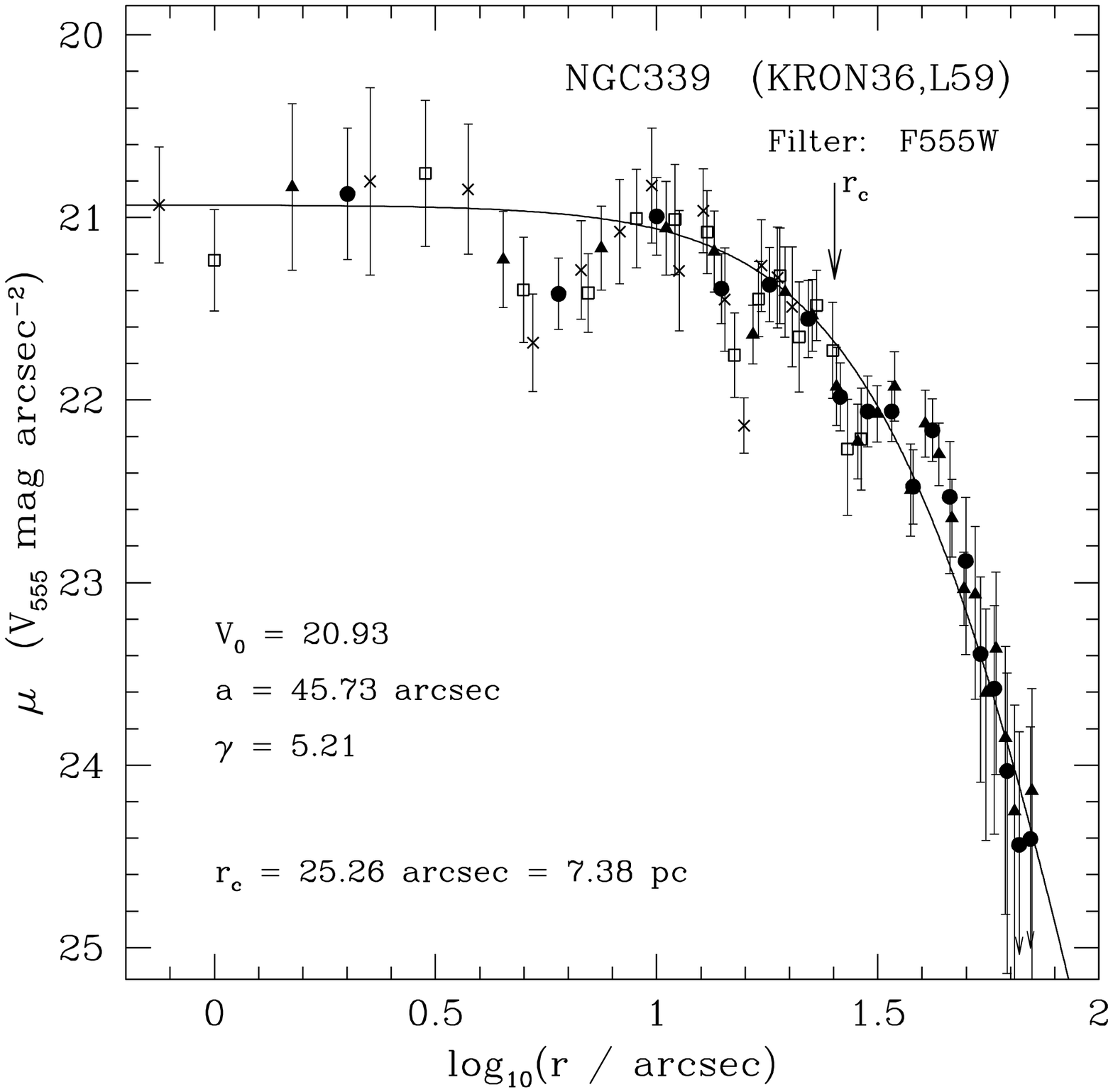}
\includegraphics[width=87mm,height=79.5mm]{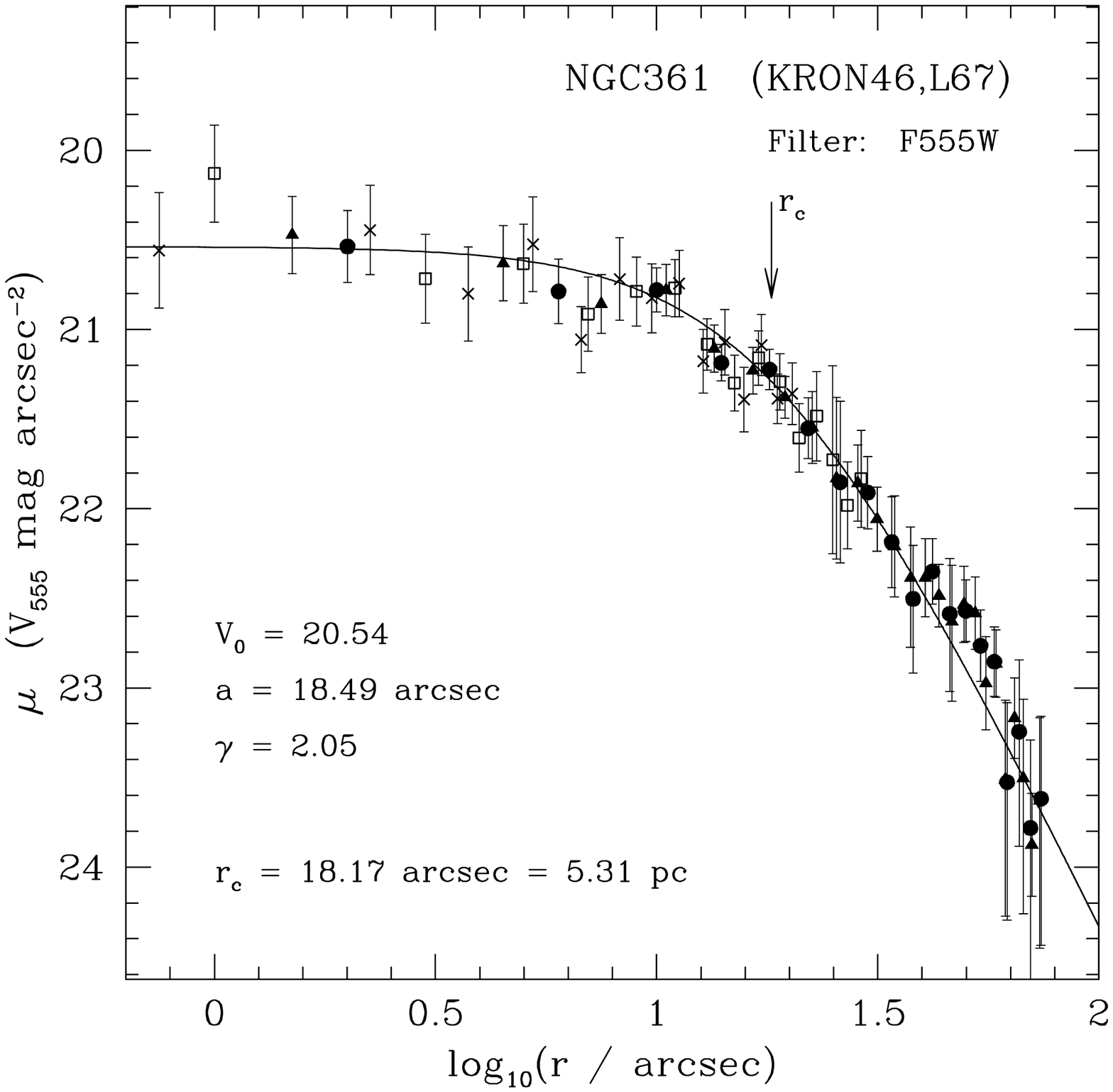}
\end{minipage}
\end{figure*}

We have located only three published studies of SMC cluster profiles,
all by the same authors -- Kontizas, Danezis \& Kontizas 
\shortcite{kdk} (20 clusters); Kontizas \& Kontizas \shortcite{kk}
(23 clusters); and Kontizas, Theodossiou \& Kontizas \shortcite{ktk}
(24 clusters). The SMC cluster profiles presented in these papers are 
in fact density profiles from number counts off UK Schmidt plates, and 
so are not strictly comparable with our results. Nonetheless, we 
proceed for completeness. All ten of the clusters from our present 
sample are included in these works -- five from the first paper listed
above, and five from the second. The values published in these papers are 
summarized in Table \ref{comp}. For the purposes of comparison, we 
convert the published core radii to our distance scale\footnote{i.e., 
from the authors' adopted distance modulus of 19.2 to ours of 18.9}. 

When the two sets of measurements are compared, we 
find significant discrepancies between them -- nonetheless, these are 
expected and can be mostly accounted for. There appears to be both a 
systematic difference and large random differences present. Eight of 
our measured core radii are larger than the corresponding published 
radii; this systematic under-estimation is expected and is in fact 
predicted by 
the authors \cite{kdk,kk}. However, the degree of under-estimation is 
apparently not systematic, with our core radii ranging from $1.3$ to 
$\sim 12$ times larger than the published radii, suggesting significant 
random effects. Again, these errors might be expected, given the
details of the photographic plate measurements. The authors report that 
their core radius estimates are derived from King models fitted to the 
density profiles, which yield $r_{t}$ and the concentration $c$, with
errors of $\sim 15$ per cent in $r_{t}$. A similar random error in $c$
could cause uncertainties of the order of $\pm 60$ per cent or more in 
derived values of $r_{c}$, consistent with the difficulties associated 
with counting stars in crowded regions on photographic plates, and the 
resolution of the profiles (most do not extend within $\sim 7$ pc). Such
uncertainties would account for most of the observed scatter, and we 
therefore do not find any cause for alarm in the discrepancies we have 
reported, especially given the demonstrated success of our reduction 
procedure for the LMC cluster sample in Paper I.

Unlike for the LMC clusters, we do not observe any evidence for
double or post core-collapse (PCC) clusters in the present sample. The lack of
PCC clusters is not surprising -- it is known that NGC 121,
the oldest documented cluster in the SMC, is some $2-3$ Gyr younger than
the oldest LMC and Milky Way globular clusters\cite{mighell,shara}. It is 
therefore likely that no SMC clusters are dynamically old enough to have entered 
PCC evolution. The lack of double clusters, which are observed in abundance
in the SMC (see e.g., the catalogue by de Oliveira et al. 
\shortcite{deol} and the references therein), can be ascribed to the
small size of our sample. None of the clusters in the sample
is present in the aforementioned catalogue. We note, however, the 
presence of a bump in the profile of NGC 152, and a bump at lower
significance in the profile of NGC 176. These bumps are reminiscent
of those observed for the LMC clusters NGC 2213 and NGC 2153 (see
Paper I), albeit not as prominent. Just as with the LMC clusters, the
causes of the bumps are not yet clear and it is uncertain as to
whether any physical significance can be attached to them. The images 
of these clusters show that there are certainly no evident sub clusters 
like NGC 1850B in the LMC.

\subsection{Luminosity and mass estimates}
\begin{table*}
\begin{minipage}{175mm}
\caption{Luminosity and mass estimates calculated using the structural parameters from the best fitting EFF profiles.}
\begin{tabular}{@{}lccccccccc}
\hline \hline
Cluster & $\log \mu_{0}$$^{a}$ & Adopted & Adopted & $\log j_{0}$ & $\log L_{\infty}$ & $\log L_{m}$ & $\log \rho_{0}$ & $\log M_{\infty}$ & $\log M_{m}$ \vspace{0.5mm} \\
 & $(L_{\odot}\,{\rmn pc}^{-2})$ & [Fe/H] & $M/L_V$ & $(L_{\odot}\,{\rmn pc}^{-3})$ & $(L_{\odot})$ & $(L_{\odot})$ & $(M_{\odot}\,{\rmn pc}^{-3})$ & $(M_{\odot})$ & $(M_{\odot})$ \\
\hline
NGC121 & $3.23 \pm 0.02$ & $-1.65$ & $2.74$ & $2.47 \pm 0.05$ & $5.12 \pm 0.10$ & $5.06 \pm 0.08$ & $2.91 \pm 0.05$ & $5.55 \pm 0.10$ & $5.50 \pm 0.08$ \vspace{0.2mm} \\
NGC152 & $2.23 \pm 0.02$ & $-0.64$ & $0.63$ & $1.17^{+0.13}_{-0.12}$ & $5.73^{+0.52}_{-0.85}$ & $4.69^{+0.18}_{-0.21}$ & $0.97^{+0.13}_{-0.12}$ & $5.53^{+0.52}_{-0.85}$ & $4.49^{+0.18}_{-0.21}$ \vspace{0.2mm} \\
NGC176 & $2.49 \pm 0.03$ & $-0.64$ & $0.23$ & $1.79 \pm 0.09$ & $4.43^{+0.22}_{-0.21}$ & $4.27^{+0.14}_{-0.15}$ & $1.15 \pm 0.09$ & $3.79^{+0.22}_{-0.21}$ & $3.63^{+0.14}_{-0.15}$ \vspace{0.2mm} \\
NGC330 & $3.61 \pm 0.02$ & $-0.64$ & $0.09$ & $2.89 \pm 0.07$ & $5.63^{+0.20}_{-0.18}$ & $5.46^{+0.12}_{-0.13}$ & $1.84 \pm 0.07$ & $4.58^{+0.20}_{-0.18}$ & $4.41^{+0.12}_{-0.13}$ \vspace{0.2mm} \\
NGC339 & $2.26 \pm 0.02$ & $-1.65$ & $1.66$ & $1.07 \pm 0.13$ & $4.80^{+0.29}_{-0.26}$ & $4.74^{+0.19}_{-0.22}$ & $1.29 \pm 0.13$ & $5.02^{+0.29}_{-0.26}$ & $4.96^{+0.19}_{-0.22}$ \vspace{0.2mm} \\
NGC361 & $2.41 \pm 0.02$ & $-1.65$ & $2.03$ & $1.39 \pm 0.08$ & $5.98^{+0.46}_{-0.66}$ & $4.82^{+0.11}_{-0.12}$ & $1.70 \pm 0.08$ & $6.29^{+0.46}_{-0.66}$ & $5.13^{+0.11}_{-0.12}$ \vspace{0.2mm} \\
NGC411 & $2.91 \pm 0.02$ & $-0.64$ & $0.63$ & $2.14 \pm 0.05$ & $4.93^{+0.14}_{-0.13}$ & $4.80 \pm 0.10$ & $1.94 \pm 0.05$ & $4.73^{+0.14}_{-0.13}$ & $4.60 \pm 0.10$ \vspace{0.2mm} \\
NGC416 & $3.20 \pm 0.02$ & $-1.65$ & $1.79$ & $2.43 \pm 0.08$ & $5.02^{+0.18}_{-0.17}$ & $4.99 \pm 0.15$ & $2.68 \pm 0.08$ & $5.27^{+0.18}_{-0.17}$ & $5.24 \pm 0.15$ \vspace{0.2mm} \\
NGC458 & $3.03 \pm 0.03$ & $-0.33$ & $0.25$ & $2.21 \pm 0.12$ & $5.03^{+0.29}_{-0.27}$ & $4.96^{+0.21}_{-0.23}$ & $1.61 \pm 0.12$ & $4.43^{+0.29}_{-0.27}$ & $4.36^{+0.21}_{-0.23}$ \vspace{0.2mm} \\
KRON3 & $2.55 \pm 0.01$ & $-1.65$ & $1.60$ & $1.46 \pm 0.05$ & $5.14 \pm 0.12$ & $4.97^{+0.07}_{-0.08}$ & $1.66 \pm 0.05$ & $5.35 \pm 0.12$ & $5.17^{+0.07}_{-0.08}$ \vspace{0.2mm} \\
\hline
\end{tabular}
\medskip
\\
$^{a}$ Corrected for reddening using $E(B-V) = 0.05$ (see text). \\
\label{luminmass}
\end{minipage}
\end{table*}

We can use the structural parameters obtained from the surface brightness
profiles to estimate luminosities and masses for each cluster, and the
procedure for this is described at length in Paper I. We showed 
that for a cluster with a profile given by Eq. 
\ref{ep}, the asymptotic luminosity is given by:
\begin{equation}
L_{\infty} = \frac{2 \pi \mu_{0} a^{2}}{\gamma - 2}
\label{linf}
\end{equation}
provided $\gamma > 2$. In cases where $\gamma \approx 2$, this value may
become unreasonably large and the limit $r\rightarrow\infty$ may not be
justified. We therefore provided a lower-limit estimate, $L_{m}$, which 
is the enclosed luminosity within a cylinder of radius $r_{m}$ along the line
of sight, where $r_m$ is the maximum radial extent measured for a given profile 
(see Table \ref{params}):
\begin{equation}
L_m = \frac{2 \pi \mu_{0}}{\gamma - 2} \left( a^2 - a^\gamma (a^2 + r_m^2 )^{-\frac{(\gamma - 2)}{2}} \right)
\label{lrmax}
\end{equation}
We obtain the central surface brightness $\mu_{0}$ in units of 
$L_{\odot}\,{\rmn pc}^{-2}$ by using the relation from Paper I, that:
\[
\log \mu_{0} = 0.4(V_{555}^{\odot} - \mu_{555}(0) + DM
\]
\begin{equation}
\hspace{15mm} + 3.1E(B-V)) + \log(3.423^{2})\,\, L_{\odot} \,{\rmn pc}^{-2}
\label{convlsun}
\end{equation}
It was shown in Paper I that $V_{555}^{\odot} = +4.85$ is an appropriate
value for the F555W absolute solar magnitude. We assume a 
uniform reddening of $E(B-V) = 0.05$, which is a reasonable approximation
in the direction of the SMC (see e.g., the values given in Table 2 of
Crowl et al. \shortcite{crowl}), and an SMC distance modulus of 18.9,
which is where the factor $3.423$ arcsec pc$^{-1}$ arises from. 

Estimates for the masses $M_{\infty}$ and $M_{m}$, and the central
density $\rho_{0}$, which correspond to $L_{\infty}$, $L_{m}$ and $j_{0}$
respectively, are obtained by multiplying the appropriate equation
by the mass to light ratio $M/L_V$ for the cluster in question. 
To obtain estimates of these ratios, we use the evolutionary
synthesis code of Fioc \& Rocca-Volmerange \shortcite{pegase} 
(PEGASE v2.0, 1999), which determines the integrated properties 
of a synthetic stellar population as a function of time. We adopted
the simplest possible model - a population of stars formed simultaneously
in one initial burst and with the same metallicity - presumably a fairly 
good approximation to the formation of a rich stellar cluster. For the
initial mass function (IMF), we used that of Kroupa, Tout \& Gilmore 
\shortcite{ktg} over the mass range $0.1$ to $120 M_{\odot}$.
There are four available abundances for these calculations 
([Fe/H] $\approx -2.25$, $-1.65$, $-0.64$, and $-0.33$) and we adopted 
the most suitable one for a given cluster based on the literature 
estimates in Table \ref{ages}. As shown in Paper I, these
calculations are relatively insensitive to the chosen metallicity --
in fact it is sufficient to determine whether a cluster is metal rich
([Fe/H] $\approx -0.33$) or metal poor ([Fe/H] $\approx -2.25$) --
and we are therefore confident in using even those metallicities from
Table \ref{ages} which are not particularly well constrained.

The mass, luminosity and density values calculated for the SMC 
cluster sample are presented in Table \ref{luminmass}. Just as for the
LMC sample, the indicated errors reflect the random uncertainties due
to the parameters $(\mu_{0},\gamma,a)$ and do not include the systematic
uncertainties which might be introduced through the mass to light 
ratios, or by the observation techniques (such as the loss of 
luminosity/mass due to saturated stars). Such systematics are not
expected to be unreasonably large -- as discussed in Paper I, they should
be within the random errors.

\section{The core radius vs. age relationship}
\label{discussion}
In Paper I, evidence for a trend in core radius with age for the LMC
cluster system was presented -- namely that the spread in cluster core
radius increases with increasing age. This trend was discussed
in detail, and the argument made that the relationship represents true 
physical evolution in these clusters. One of the first things we would 
like to ask for the present sample is whether such a relationship is 
also present for the SMC cluster system, and if so whether it is 
noticeably different in any way from that observed in the LMC.

\begin{figure}
\includegraphics[width=0.5\textwidth]{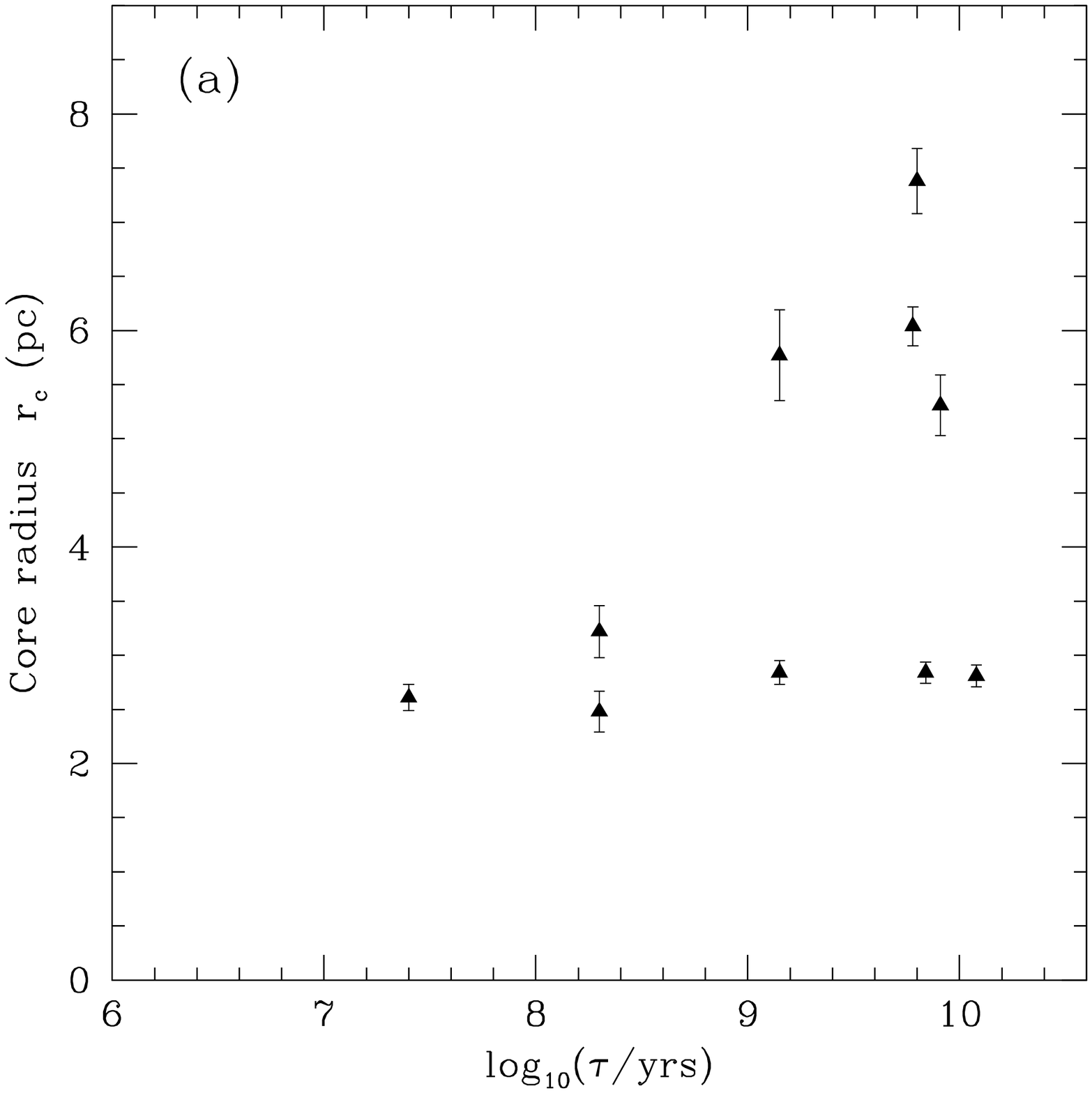}
\includegraphics[width=0.5\textwidth]{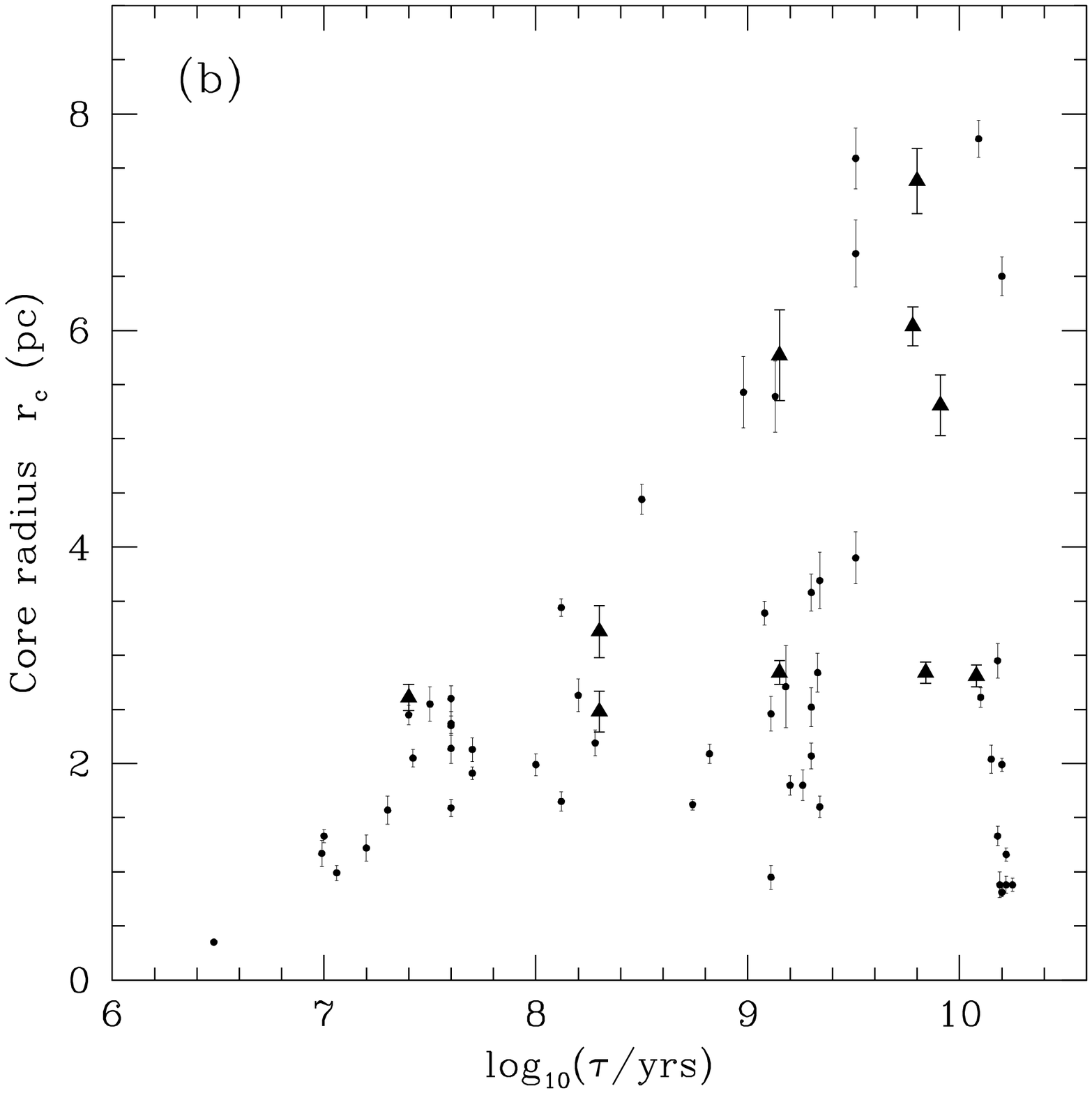}
\caption{Core radius vs. age for {\em (a)} all ten SMC clusters and {\em (b)} SMC clusters (solid large triangles) and LMC clusters (small solid circles) together. Data for the SMC cluster ages are from Table 2 and core radii from Table 3; all LMC cluster data is from Paper I.}
\label{smcagecore}
\end{figure}

We have plotted core radius against age for the ten clusters, in Figure
\ref{smcagecore}(a). Even though our sample is small and 
therefore incomplete, it immediately seems that the same
trend in core radius exists for the SMC cluster system. This is the
first time that this has been shown for SMC clusters. The youngest
clusters have relatively compact cores, whereas the older clusters
can have either compact cores or much larger cores. The similarity of
the SMC cluster trend to that for the LMC sample is evident when the
data for both systems are over-plotted (Figure 
\ref{smcagecore}(b)). Aside from the fact that four of the SMC clusters
fall in the LMC cluster age gap, no significant differences 
between the two relationships can be claimed, at least within 
the limits of the current data. Both plots show a bifurcation at 
approximately several$\times 10^{8}$ yr, with the majority of clusters
following the lower sequence in the diagram,
but with some clusters moving to the upper right. The two pairs of
sequences fall nicely on top of each other. If anything, the SMC
clusters on the lower sequence have slightly larger core radii
on average than the LMC lower sequence clusters; however, this could be due 
to uncertainties in both the SMC distance modulus and line-of-sight spatial 
depth. It is tempting to suggest that a larger fraction of SMC 
clusters than LMC clusters evolve to the upper sequence, but given the small 
size of the SMC sample, we cannot claim this without additional data.

\begin{figure}
\includegraphics[width=0.5\textwidth]{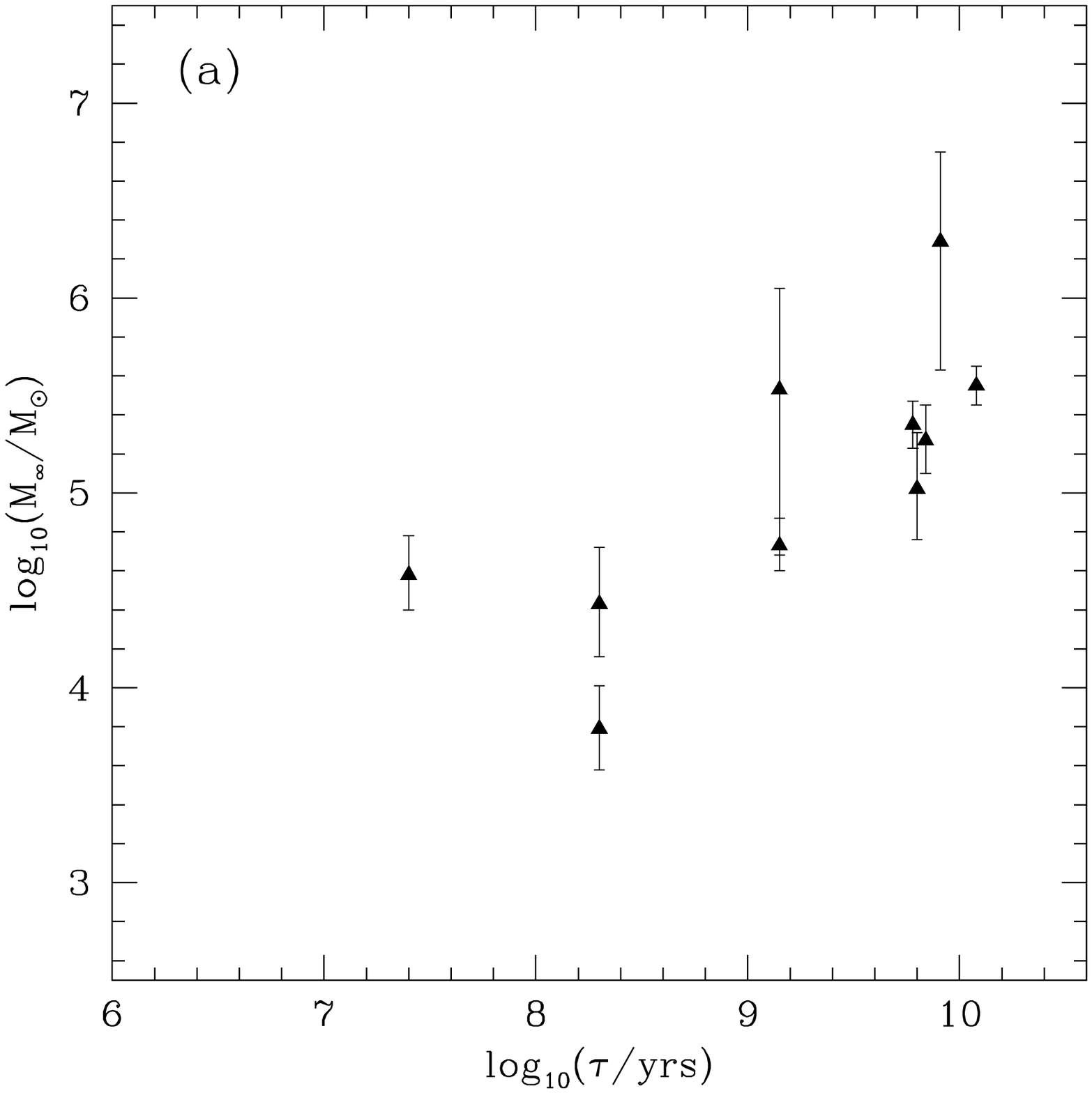}
\includegraphics[width=0.5\textwidth]{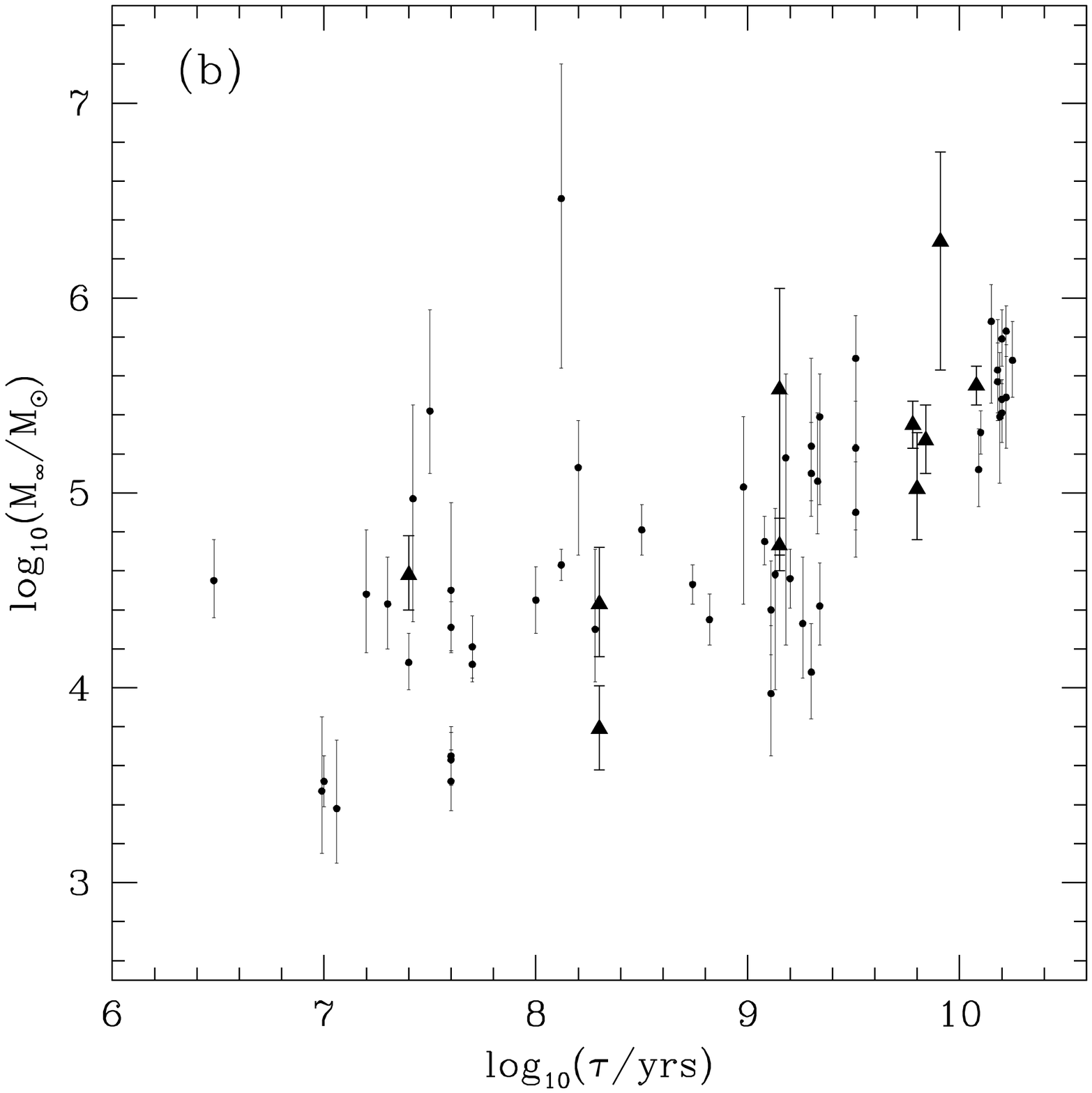}
\caption{Asymptotic mass vs. age for {\em (a)} all ten SMC clusters and {\em (b)} SMC clusters (solid large triangles) and LMC clusters (small solid circles) together. Data for the SMC cluster ages are from Table 2 and masses from Table 5; all LMC cluster data is from Paper I.}
\label{smcagemass}
\end{figure}

\begin{figure}
\includegraphics[width=0.5\textwidth]{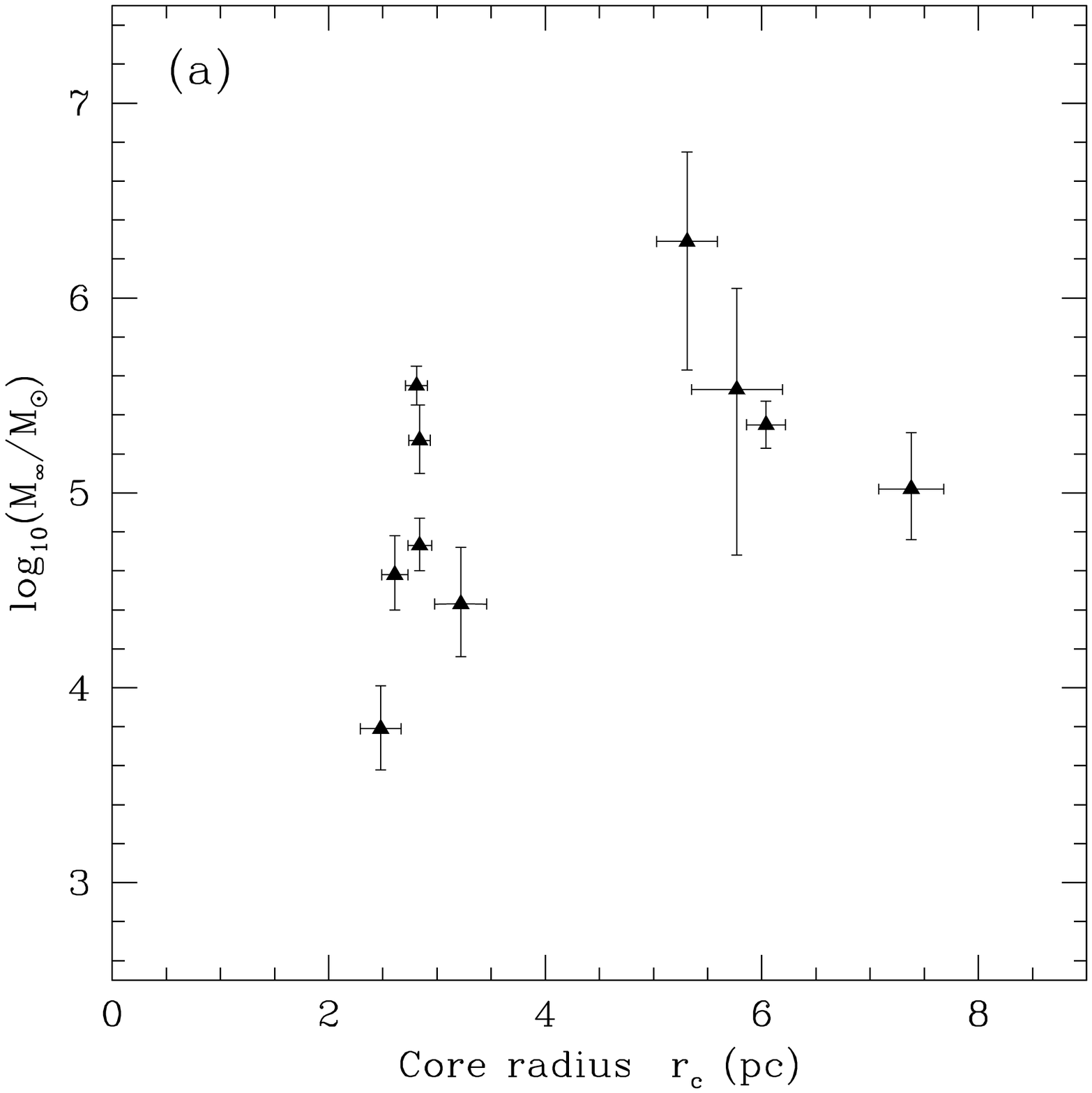}
\includegraphics[width=0.5\textwidth]{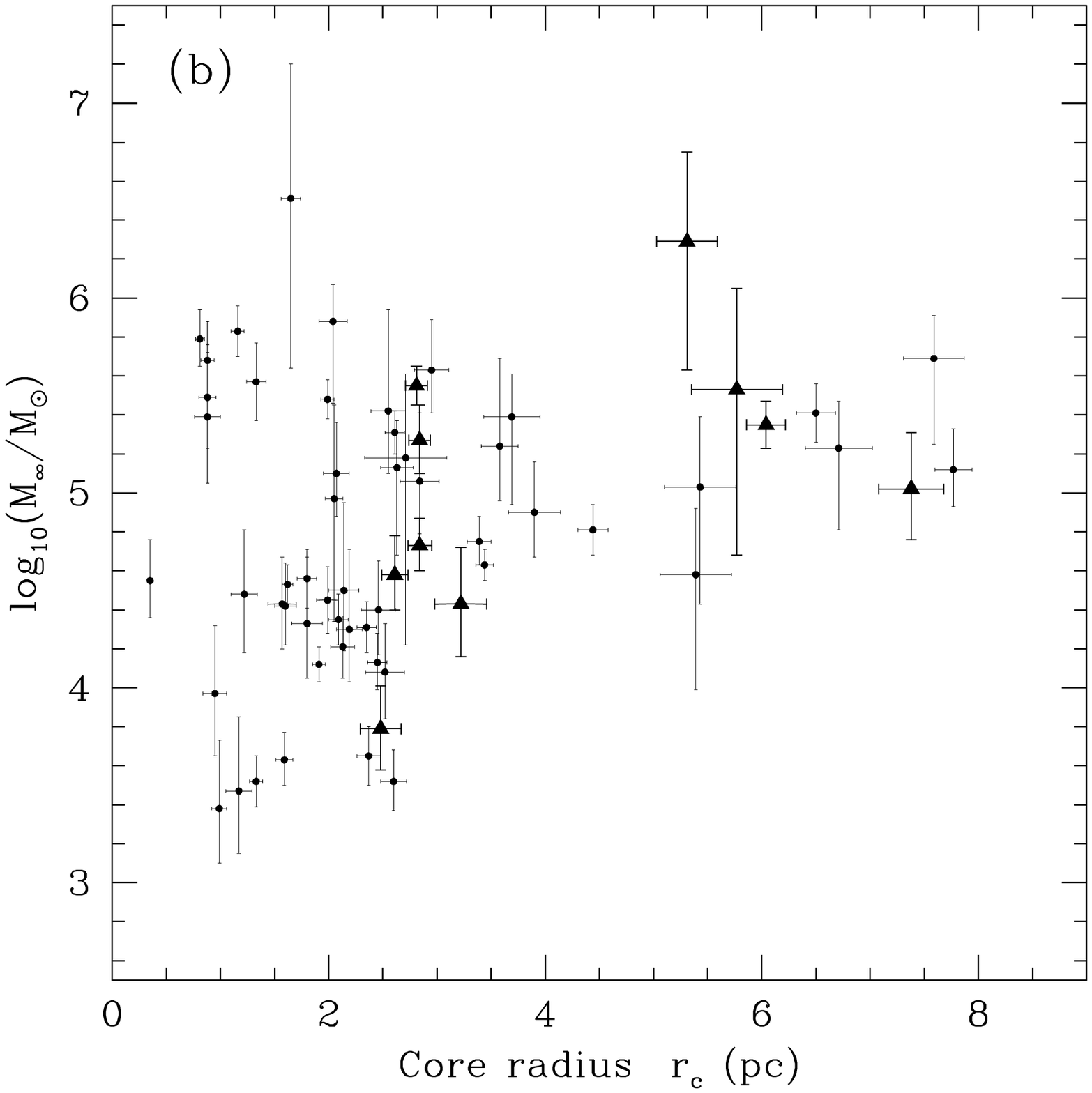}
\caption{Asymptotic mass vs. core radius for {\em (a)} all ten SMC clusters and {\em (b)} SMC clusters (solid large triangles) and LMC clusters (small solid circles) together. Data for the SMC cluster core radii are from Table 3 and masses from Table 5; all LMC cluster data is from Paper I.}
\label{smccoremass}
\end{figure}

In Paper I, we used two diagnostic plots -- $M_{\infty}$ vs. $\log\tau$ 
and $M_{\infty}$ vs. $r_{c}$ -- to show that the radius-age 
relationship is not due to a correlation between mass and age, and
mass and core radius. We can use the current data both to demonstrate
this for the SMC clusters, and also to search for any systematic 
differences which might be present between the two cluster systems and 
which might help shed light on the core radius vs. age trend. The two 
diagnostic plots are presented in Fig. \ref{smcagemass}(a) and Fig.
\ref{smccoremass}(a) respectively. The LMC data is added to these
diagrams in Fig. \ref{smcagemass}(b) and Fig. \ref{smccoremass}(b). 

Fig. \ref{smcagemass}(a) shows the possibility of a slight correlation of
SMC cluster mass with age; however this apparent correlation is probably
due to the incompleteness of the sample, particularly 
for clusters younger than $10^{9}$ yr. When compared with the LMC
sample, which showed no such correlation, the agreement is close. 
The lowest mass SMC cluster is not
of unusually low mass; similarly, the highest mass cluster is not
overly massive. In fact, without the addition of more SMC clusters to
the sample, the plots for the two systems cannot be considered to show
any significant differences. The same result applies for the plot of 
$M_{\infty}$ against $r_{c}$. No correlation is evident from the SMC 
data alone, and there are no significant differences between the SMC and
LMC data -- in fact, the agreement is extremely close. This
further strengthens both the conclusion that there is little difference
between the core radius vs. age relationship for each system, and 
that this relationship represents real evolution in the structure of
clusters as they grow older. Furthermore, if the suggestion that
a higher fraction of SMC than LMC clusters have expanded cores {\em does}
hold true given more data, this is unlikely to be due to significantly different 
cluster mass distributions between the two systems.

The apparent close agreement between the structural parameters of 
the LMC and SMC clusters is potentially significant 
given that the two systems are known to exhibit strong differences in 
other aspects.  The LMC apparently started forming globular clusters
some $2-3$ Gyr before the SMC, as evidenced by the age ($11.9\pm1.3$ Gyr) of
the oldest known SMC cluster, NGC 121 \cite{mighell}, compared with
the ages ($\sim 16\pm3$ Gyr) of the oldest LMC clusters (Olsen et al. 
\shortcite{olsen}; see also Shara et al. \shortcite{shara}). 
Furthermore, the LMC seems to have had a long quiescent period in cluster 
formation, with only one cluster (ESO 121-SC03) definitely within the so 
called ``age gap'' between $\sim 3 - 12$ Gyr (see e.g, Rich, Shara \& Zurek 
\shortcite{rsz} and references therein). The SMC on the other hand, 
continued forming clusters during this period, possessing at least seven
with ages lying in the age gap range (see e.g., Mighell et al. 
\shortcite{mighell}). The age-metallicity 
relationships for the two systems are correspondingly different -- 
the LMC possesses an abundance gap which matches the age gap
(again see e.g., Rich et al. \shortcite{rsz})
but appears to have undergone considerable chemical enrichment during
this period \cite{mighell}. For a detailed discussion of this, we 
refer the reader to the study of Gilmore \& Wyse \shortcite{gilw} who
investigate the chemical evolution of systems which undergo bursts of 
star formation, with particular application to the LMC. In contrast with
the LMC, the SMC has plenty of intermediate 
metallicity clusters (see e.g., Da Costa \& Hatzidimitriou 
\shortcite{gdc}) and does not seem to have undergone such a steep
chemical enrichment in the age gap period. The data of Mighell et al. 
suggest that a bursting model of star formation 
might best fit the SMC age-metallicity relation, with a 2 Gyr peak
of formation around $11$ Gyr ago, then a lower but constant 
formation rate until $\sim 2$ Gyr ago. Da Costa \& Hatzidimitriou
suggest a closed box model of chemical enrichment best
fits their (slightly different) spectroscopic metallicity determinations.

It is intriguing that two systems with such evidently disparate cluster
formation histories should nonetheless possess cluster systems which
appear to have followed such similar evolutionary paths. This again
reinforces the idea that the trend in core radius with age
observed for both systems is purely a result of some aspect of
cluster evolution; an aspect which should operate similarly in each of
the two Magellanic Clouds. The need for additional data is clear and
paramount, and the acquisition of this data is a crucial part of the
future development of this work. $N$-body simulations are also underway
(e.g., Wilkinson et al., in prep.),
exploring how different stellar populations, and external
influences such as those described in Paper I can affect the evolution
of cluster cores.

\section{Summary and Conclusions}
In a follow-up to our recent work detailing the structures of a large
sample of LMC clusters (Paper I), we have obtained a similar archival 
{\em HST}  snapshot data set for a sample of ten SMC clusters. We have 
constructed surface brightness profiles for these clusters, and obtained 
measurements of their structural parameters, following a method exactly 
similar to that applied to the LMC sample. Luminosities, masses and 
central densities have also been estimated for the SMC sample. These 
data, along with the surface brightness profiles, are available on-line 
at {\em http://www.ast.cam.ac.uk/STELLARPOPS/SMC\_clusters/}. Unlike for 
the LMC sample, we do not see any evidence for post core-collapse 
clusters in our sample, but this is not unexpected. Similarly, we do not 
see any compelling evidence from the surface brightness profiles for 
double clusters in our sample, although a couple of profiles show 
bumps similar to those observed for several LMC clusters.

We have used our core radius measurements for the SMC sample to
investigate further the core radius vs. age relationship, which was
described in detail in Paper I. Although compromised somewhat by the
small sample size, our analysis shows that the SMC clusters apparently 
follow a very similar relationship to the LMC clusters, with some
clusters maintaining small cores throughout their lives, but with others
developing much enlarged cores. It is possible that a higher percentage 
of rich SMC clusters than rich LMC clusters develop such expanded cores.
Additional data, both observational and computational, is 
required to further explore this relationship and the physical processes 
involved.

\section*{Acknowledgments}
ADM would like to acknowledge the support of a Trinity College ERS 
grant and a British government ORS award.
This paper is based on observations made with the NASA/ESA 
{\em Hubble Space Telescope}, obtained from the data archive at the 
Space Telescope Institute. STScI is operated by the association of 
Universities for Research in Astronomy, Inc. under the NASA contract 
NAS 5-26555.

%\bibitem[\protect\citename{Fischer et al.\ }1998]{fischer}
%  Fischer P., Pryor C., Murray S., Mateo M., Richtler T., 1998,
%  AJ, 115, 592
%\bibitem[\protect\citename{Ripepi et al.\ }1999]{ripepi}
%  Ripepi V., Brocato E., Castellani V., 1999, A\&A, 351, 526

% End of paper here

%%%%%%%%%%%%%%%%%%%%%%%%%%%%%%%%%%%%

\bsp % ``This paper has been produced using the ...''

\label{lastpage}

\end{document}